\documentclass{aa}  
\usepackage{graphicx}
\usepackage[dvipsnames]{xcolor}

\usepackage{txfonts}
\usepackage{natbib}
\usepackage{xcolor}
\usepackage{url}
\usepackage{multicol}
\usepackage{multirow}
\usepackage[utf8]{inputenc}
\usepackage{amsmath}
\usepackage{nccmath}
\usepackage{amsfonts}
\usepackage{amssymb}
\usepackage{graphicx}
\usepackage{comment}
\usepackage{caption}
\usepackage{subcaption}
\usepackage{soul}
\usepackage[normalem]{ulem}
\usepackage[colorlinks=true,linkcolor=blue,citecolor=blue, urlcolor=blue]{hyperref}
\usepackage{siunitx}

\usepackage{lscape}
\usepackage{longtable}
\usepackage{gensymb}

\newcommand\soutblue{\bgroup\markoverwith{\textcolor{blue}{\rule[.5ex]{2pt}{1.5pt}}}\ULon}

\begin{document}


\title{
Consequences of the lack of azimuthal freedom in the modeling of lensing galaxies}
\author{Lyne Van de Vyvere\inst{1}\fnmsep \thanks{\email{lyne.vandevyvere@uliege.be}}
          \and
          Dominique Sluse\inst{1}
          \and
          Matthew R. Gomer\inst{1}
          \and
          Sampath Mukherjee\inst{1}
          }

\institute{STAR Institute, Quartier Agora - All\'ee du six Ao\^ut, 19c B-4000 Li\`ege, Belgium }

   \date{Received DD--MM--YYYY; accepted DD--MM--YYYY }

  \abstract
   {Massive elliptical galaxies can display structures that deviate from a pure elliptical shape, such as a twist of the principal axis or variations in the axis ratio with galactocentric distance. Although satisfactory lens modeling is generally achieved without accounting for these azimuthal structures, the question about their impact on inferred lens parameters remains, in particular, on time delays as they are used in time-delay cosmography. This paper aims at characterizing these effects and quantifying their impact considering realistic amplitudes of the variations. 
   We achieved this goal by creating mock lensing galaxies with morphologies based on two data sets: observational data of local elliptical galaxies, and hydrodynamical simulations of elliptical galaxies at a typical lens redshift. We then simulated images of the lensing systems with space-based data quality and modeled them in a standard way to assess the impact of a lack of azimuthal freedom in the lens model. We find that twists in lensing galaxies are easily absorbed in homoeidal lens models by a change in orientation of the lens up to 10\degree~with respect to the reference orientation at the Einstein radius, and of the shear by up to 20\degree~with respect to the input shear orientation. The ellipticity gradients, on the other hand, can introduce a substantial amount of shear that may impact the radial mass model and consequently bias $H_0$, up to 10 $\text{km}\,\text{s}^{-1}\,\text{Mpc}^{-1}$. However, we find that light is a good tracer of azimuthal structures, meaning that direct imaging should be capable of diagnosing their presence. This in turn implies that such a large bias is unlikely to be unaccounted for in standard modeling practices. Furthermore, the overall impact of twists and ellipticity gradients averages out at a population level. For the galaxy populations we considered, the cosmological inference remains unbiased.
   }
   
\keywords{Gravitational lensing: strong, Galaxies: elliptical and lenticular, cD, Methods: numerical, Cosmological parameters}

\maketitle

\section{Introduction}

In the past decades, strong gravitational lensing analysis has been a competitive method used to constrain the Hubble constant $H_0$. Along with other methods \citep[e.g.,][]{Abbott2018,Riess_cepheid,Freedman2020_tip_red,Pesce2020_megamaser,BAO_philcox2020,PlanckVI,Schombert2020_tullyfisher,HolicowXIII,Blakeslee2021_IRsb}, the key challenge has been to reach an increasingly precise value of $H_0$. Increasing the precision of each measurement led to a difference between the early-time probes, such as cosmic microwave background, and late-time probes, such as cepheid and supernova methods \citep{Verde2019}. Strong gravitational lensing has the advantage of being independent of any distance ladder while being a late-time probe. This technique is based on the phenomenon that a massive galaxy bends the light arising from a variable source, such as a quasar, and thus produces up to several images of this source \citep{Refsdal1964}. Modeling the mass of the foreground galaxy allows constraining the effect of the galaxy (displacement and elongation) on the light rays coming from the source. In combination with the arrival time difference between the different images of the source, the Hubble constant can be constrained assuming a given cosmological model. Other factors, such as the environment surface mass density, the velocity dispersion of the lensing galaxy, and time-delay microlensing effects, also help characterizing $H_0$ with precision \citep[e.g.,][]{Chen2018,Sluse2019,Tihhonova2020,TDCOSMOII,Donnan2021,Liao2021,TDCOSMOVIII}.

While a gravitational lensing analysis requires many ingredients to perform with the best precision, we focus here on the role of the mass model of the lensing galaxy. Most lensing galaxies are massive ellipticals. Those early-type galaxies are mainly modeled by either a power-law ellipsoid or by a combination of a dark and baryonic ellipsoidal components in strong-lensing studies. From the point of view of the lensing mass distribution, the assumption about the radial mass profile has been explored extensively \citep{SS2013, SS2014, Xu2016, Kochanek2020, TDCOSMOIV,Kochanek2021}, while the azimuthal mass profile has been explored less frequently. Recently, \cite{multipole} specifically addressed the question of the impact of boxyness or diskyness, that is, the octupolar moment in addition to the usual quadrupolar (elliptical) lens, on the consmographic inference if only quadrupolar-mass models are used. \cite{Cao2022} studied the influence of the shape of lensing galaxies under the elliptical power-law (EPL) model assumption in more detail. They used simulations of mock images whose lensing galaxy mass morphology was based on observed SDSS-MaNGA\footnote{SDSS stands for Sloan Digital Sky Survey, and MaNGA for Mapping Nearby Galaxies at APO, with APO standing for the Apache Point Obseratory} stellar dynamics data, and cautioned about too simple (i.e., EPL) lens-mass models. \cite{Kochanek2021} also cautioned about the possible effect of angular variations in the lensing galaxy on the Hubble constant when the model lacks degrees of freedom in the azimuthal direction. These setups can allow clear but incorrect likelihood distinctions between different radial mass profiles that sometimes lead to an apparently precise but biased $H_0$ determination.

To quantify the role of azimuthal structure on gravitational lensing modeling inference, it is necessary to understand how strong the azimuthal perturbations are in both light and total mass profile, and in particular, in the inner regions of galaxies, which prevails in the lensing phenomenon. 
The galaxy formation processes in Lambda cold dark matter ($\Lambda$CDM) universe considers small fluctuations in dark matter density that are seeded during inflation, to grow over time through mergers with other halos and/or slow accretion, forming the dark matter halos around present-day galaxies \citep{davis1985,frenk1988}. A number of early-type galaxy formation scenarios have been proposed over the past five decades: (i) monolithic collapse, (ii) major mergers between galaxies, leading to larger galaxies, and (iii) minor mergers, in which the cores of massive galaxies extend, but not much more mass is added. The inflow of gas through accretion and mergers in to the centers of dark matter halos can modify their density profiles and shapes through adiabatic contraction (or expansion). This can be further complicated by possible outflows induced by feedback from stellar winds, SN explosions, UV radiation, and AGN feedback, for instance. Because a complete analytic theory of baryonic physics is lacking, cosmological hydrodynamic simulations that include many physical processes have emerged as the dominant tool for studying the complex nonlinear interactions taking place during galaxy formation \citep[e.g.,][]{vogel2014,Schaye2015}. The most recent hydrodynamical simulations with improved stellar and AGN feedback, for example, can reproduce the cosmic star formation history of the Universe and the galaxy stellar mass function. In summary, numerical simulations and recent large surveys have helped us to understand galaxy formation more precisely, but some questions remain open because we generally only have access to the luminous mass.

Nevertheless, morphological features of galaxies could provide important information about their formation history \citep{Hao2006,Kormendy2009,Chaware2014,Cappellari2016}. Early surface photometry measurements of large samples of galaxies have shown different estimates of the percentage of significantly twisted objects among the galaxies. The estimates range from 10\% to 60\% \citep{Lauer1985a, Lauer1985b, Michard1985, Djorgovski1986, Jedrzejewski1987, Vigroux1988}. This broad disagreement can partially be described by the different criteria that are used to select the samples, the data quality of the images,  and by the definition of the twist itself. Moreover, several factors unrelated to the intrinsic structure of the objects may be able to produce isophotal twisting, such as dust in lanes or patches, the overlap with isophotes of nearby projected companions, tidal effects due to nearby galaxies, a noncircular point spread function, and artificial trends in the background even after flat fielding \citep{Kormendy1982,Fasano1989}. Nevertheless, intrinsic azimuthal variations with the major axis of elliptical galaxies is now acknowledged. For instance, \cite{Hao2006} analyzed the shape of 847 early-type galaxies and reported a standard deviation of 8.39\degree\,for the position angle variation between 1 and 1.5 half-light radii in their sample, and a 0.051 scatter for the ellipticity variations. \cite{Kormendy2009} focused on all known elliptical galaxies in the Virgo cluster and reported the position angle, ellipticity, and brightness profiles of those galaxies, linking the properties of the galaxies with a formation history. Corroborated by other studies \citep[e.g.,][]{Krajnovic2011,Fogarty2015,Cappellari2016} and with the ingress of kinematics data, the formation history of elliptical galaxies, distributed into slow rotators and fast rotators, is becoming better understood. Slow rotator galaxies, with masses typically below $2 \times 10^{11} \text{M}_\odot$, arise from spiral galaxies, are distributed in space accordingly, and are consistent with oblate spheroid shapes. On the other hand, more massive fast rotators are present in locally denser environments and are likely to display a triaxial symmetry. With the advent of understanding the galaxy formation and the structure of early-type galaxies, further models of elliptical galaxies are expected to be able to reach an increased precision. 

Only few works have intentionally taken possible twists and ellipticity variations with radius in the modeling of strong lensing galaxies into account. \cite{Schramm1994} developed a framework to create any type of lensing galaxy profile through an ensemble of elliptical slices, but his work was not used in practical applications.  \cite{Keeton2000Q0957} used a double pseudo-Jaffe model, with different core radii, cutoff radii, position angles, and ellipticities, to account for twist and ellipticity variations in the lens mass profile of the lensing system Q0957+561. \cite{Fadely2010Q0957} refined the model of the same lensing galaxies. They explicitly used the light map of the galaxy with a modeled constant mass-to-light ratio to account for the baryonic component, using fast Fourier transform methods to convert the mass maps into the lensing quantities. For the dark matter component, either a Navaro-Frenk-White profile \citep[NFW;][]{NavarroFrenkWhite1996} or a combination of three softened power-law profiles were used. The twists and ellipticity changes of the light profile were thus explicitly used in the mass model for this specific system Q0957+561. More generally, using multiple analytical components in a mass model can inherently account for a gradient in ellipticity or position angle. In addition, nonparametric mass models \citep[e.g.,][]{Liesenborgs2009,Lefor2013,Lubini2014} naturally allow for more azimuthal freedom. However, these free-form or multicomponent models have not been used to quantify the impact of azimuthal variations, but rather naturally account for it, sometimes even affording a larger freedom than is present in galaxies.

In this paper we do not model known lens systems with mass profiles increased in azimuthal structures, but instead systematically investigate the impact that azimuthal perturbations have on lensed image morphologies by means of simulations.
Specifically, we create mock lensing galaxies with varying position angle and/or ellipticity, simulate lensing images of a background source, and model the images with standard lensing modeling methods, that is, we do not take the varying azimuthal structure into account. This experiment allows us to quantify the impact of the lack of azimuthal degree of freedom in a model of a lensing galaxy that displays twists and/or ellipticity gradients. 
This paper is complementary to \cite{multipole}, who investigated the impact of another type of azimuthal variations in lensing galaxies, that is, boxyness and diskyness.

In Sect. \ref{sect_method} we first review the method of \cite{Schramm1994} to simulate a lens mass profile displaying a varying position angle and ellipticity (Sect. \ref{sect_lens_mass_prof_theory}). We then apply it (Sect. \ref{sect_lens_mass_profile_practice}), explaining the strategy we followed: we first introduced an academical experiment, and then a more realistic one. We end our method section with the characteristics of our mock images (Sect. \ref{sect_mock_images}) and standard fitting procedure (Sect. \ref{sect_fiiting_procedure}). In Sect. \ref{sect_results} we display the results of the different experiments. Finally, we summarize our findings and conclude in Sect. \ref{sect_conclu}.

The fiducial cosmology used in our study is a flat $\Lambda$CDM cosmological model. The associated cosmological parameters are $\Omega_{\rm m}=0.3$, $\Omega_\Lambda=0.7$, and $H_0=70$ $\text{km}\,\text{s}^{-1}\,\text{Mpc}^{-1}$.

\section{Method}
\label{sect_method}

To study the effect of twists and ellipticity changes with radius on lensed images, we designed the following experiment. First, different lens mass profiles displaying twists and/or ellipticity variations were created. These mass profiles were then used to simulate mock images with HST data-quality by lensing a background source, consisting of a quasar and its host galaxy. The time delays associated with the lensed quasar images were also calculated. Afterward, the simulated frames were modeled using a power-law elliptical mass distribution (PEMD) and shear for the mass model, thus displaying neither twist nor ellipticity changes in the model. After the modeling process, we assessed the goodness of the fit and analyzed the retrieved parameters to correlate the varying input profiles with specific modeled parameters, such as $H_0$. This method is similar to the one presented in \cite{multipole}. 
As this paper and \cite{multipole} focus on different types of azimuthal variations, the lensing galaxies are built differently: this paper uses a sum of slices that radially mimic a singular isothermal ellipsoid (SIE) profile, while \cite{multipole} used an analytical SIE perturbed by a global octupolar moment. Nevertheless, the scientific procedure is the same in the two works as they both quantify how the input azimuthal perturbations in lensing galaxies impact the parameter inference with a mass model displaying a single elliptical shape.
\subsection{Lens mass profile: Theory}
\label{sect_lens_mass_prof_theory}
We used the technique developed by \cite{Schramm1994} to construct a mass distribution that allows the position angle to twist and ellipticity to vary with radius. This technique consists of approximating any mass profile by a sum of finite elliptical slices of constant surface mass density that can have different ellipticities, position angles, centers, and major-axis lengths. The superposition of the slices is a discrete approximation of the global mass profile. For each slice, the following quantities can be calculated: the lensing potential, its first derivative, that is, the deflection angles, and its second derivatives. They are then summed to compute the lensing quantities associated with the global mass profile. The lensing quantities associated with one slice are separated into two parts: those for a position inside the slice, and those for a position outside of the slice. If we consider an elliptical slice of constant surface mass density $\Sigma_0$ with a semi-major axis $a$, semi-minor axis $b$, and a position angle $\varphi$, the lensing potential at position $(x,y)$ inside a slice centered at coordinates $(0,0)$ is given by
\begin{multline}
    \Psi_{\rm in} = \frac{1}{2}\left( (1-\epsilon)(x\cos\varphi+y\sin\varphi)^2+ \right.\\\left.(1+\epsilon)(y\cos\varphi-x\sin\varphi)^2 \right) \Sigma_0  + \Sigma_0 r_{\rm E}^2 (1-\epsilon^2) \ln r_{\rm E,}^{~}
\label{pot_in}   
\end{multline}
where the elliptical parameter $\epsilon$ is $\frac{a-b}{a+b}$, and the elliptical radius $r_{\rm E}^{~}$ is equal to $\frac{a+b}{2}$. Outside the slice, the lensing potential is given by\begin{multline}
    \Psi_{\rm ext} = \Re \left[ \frac{1-\epsilon^2}{4\epsilon} \left( f^2 \ln{\left(\frac{\text{sign}(z e^{-i\varphi})ze^{-i\varphi}+\sqrt{z^2e^{-2i\varphi}-f^2}}{2}\right)} - \right. \right. \\ \left. \left. \text{sign}(z e^{-i \varphi})ze^{-i\varphi} \sqrt{z^2e^{-2i\varphi}-f^2} + z^2e^{-2i\varphi} \right) \Sigma_0 \right],
\label{pot_ext}
\end{multline}
where $z=x+iy$, $f^2= a^2-b^2$, and the sign function is defined such that
$$\text{sign}(z)= \begin{cases} +1 & \rm{if~} x>0 \rm{~~or~~} (x=0 \rm{~and~}y \geq 0) \\ -1 & \rm{otherwise.}    \end{cases}$$
The additive constant, that is, $\Sigma_0 r_{\rm E}^2 (1-\epsilon^2) \ln r_{\rm E}^{~}$,  in the $\Psi_{\rm in}$ definition was chosen such that the transition between external and internal potentials of a given slice was continuous. The deflection angle, $\alpha=\alpha_x+i \alpha_y$, inside the slice is given by
\begin{equation}
    \alpha_{\rm{in}} = (z-\epsilon \overline{z} e^{2 i \varphi}) \Sigma_0.
\label{alpha_in}
\end{equation}
The deflection angle outside the slice is
\begin{equation}
    \alpha_{\rm{ext}}=\frac{2 ab}{f^2}\left( \overline{z}e^{2i\varphi}-e^{i\varphi}\text{sign} ( \overline{z}e^{i\varphi} ) \sqrt{\overline{z}^2 e^{2i\varphi}-f^2}  \right) \Sigma_0.
\label{alpha_ext}
\end{equation}

The second derivatives of the potential, which are also the first derivatives of the deflection angle, can be calculated numerically by deriving the deflection angle. We implemented such a mass profile under the name 'ElliSLICE' in the \texttt{lenstronomy} software \footnote{\url{https://github.com/sibirrer/lenstronomy}}.

\begin{figure}
    \centering
    \includegraphics[width=0.49\textwidth]{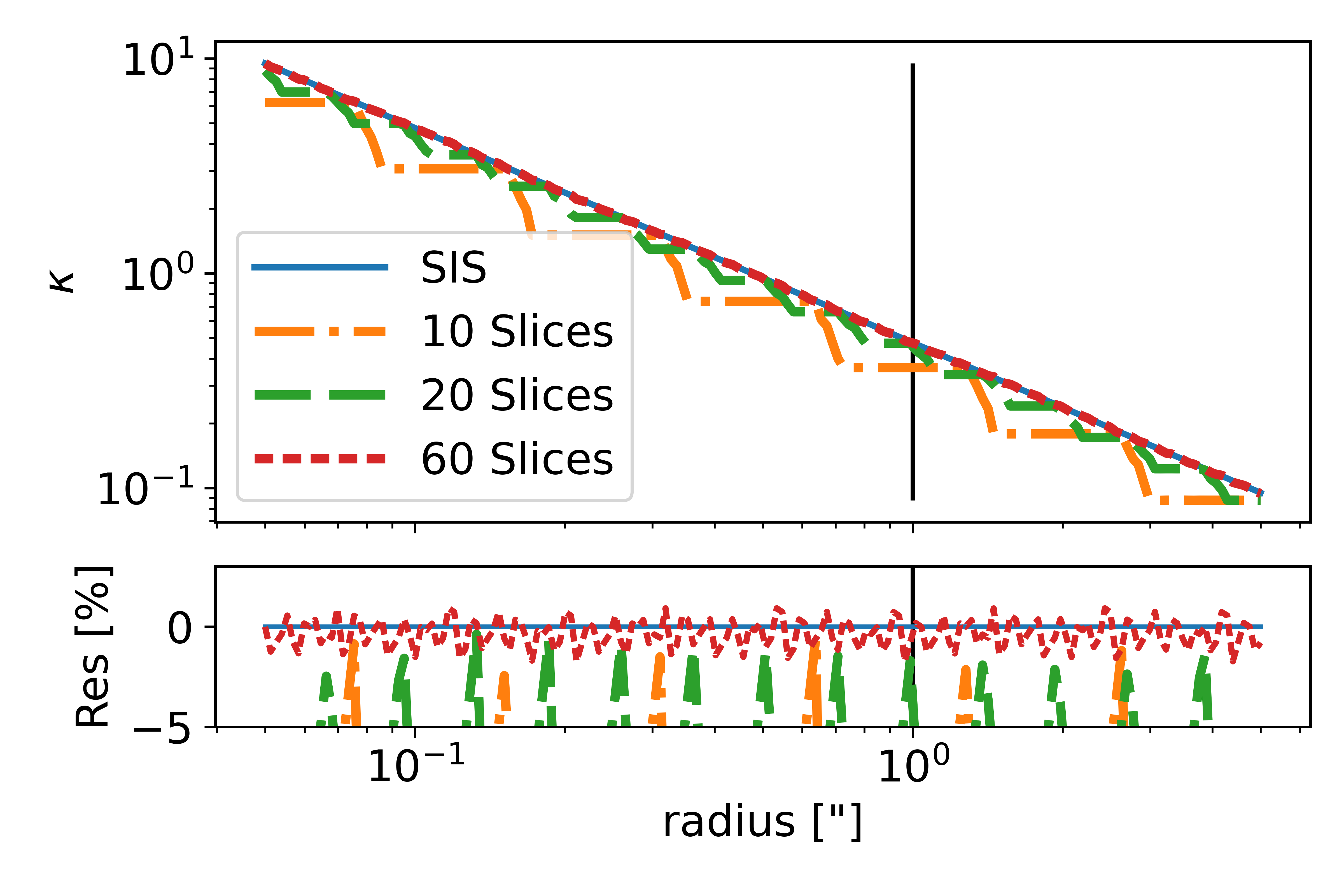}
    \caption{Accuracy of mass profiles created via slices following \cite{Schramm1994} compared to an analytical SIS. Top: Radial convergence profile of a SIS (blue line) and of 'ElliSLICE' mass models using 10, 20, or 60 slices (dash-dotted orange line, dashed green line, and densely dashed red line, respectively). Bottom: Residuals in percent between the SIS profile and profiles made of slices. The vertical black line indicates the Einstein radius.}
    \label{fig_slice_convergence}
\end{figure}

To create a realistic lens mass profile, several slices have to be superposed. The lensing quantities (i.e., lensing potential and deflection angle) associated with each slice were calculated following Eq. \ref{pot_in}--\ref{alpha_ext} , and they were then summed to constitute the lensing quantities associated with the whole profile. As the number of slices increases, the stepwise mass density approaches a smooth profile, but the computational time to calculate the lensing quantities of each slice and sum them grows as well. Approximately 60 slices in which the major axis is evenly spaced in log space are sufficient to accurately describe a realistic lensing mass profile, as shown in Fig. \ref{fig_slice_convergence} in which mass profiles created with slices are compared to a singular isothermal sphere (SIS). Sect. \ref{sect_controlled_experiment} and Fig. \ref{fig_in_out} also later show that an HST-like mock image simulated with a lensing mass profile made of 60 slices following an SIE is perfectly modeled with an SIE mass model. Typically, the last slice extends up to five times the Einstein radius. Since each slice is associated with an analytical potential and deflection angle, these simulations show no spurious artificial shear, which occurs when truncated mass maps alone are used to compute the lensing quantities. The mass profile therefore does not have to be extended up to 50 times the Einstein radius  \citep{VandeVyvere2020}.

In addition to the the mass of the lensing galaxy, a shear was also considered to mimic any tidal perturbation caused by galaxies along the line of sight or at the lens redshift. The lensing potential at position $(x,y)$ associated with a shear is the following:
\begin{equation}
\Psi_{\text{shear}}= \frac{1}{2} (\gamma_1 x^2 + 2 \gamma_2 xy - \gamma_1 y^2)
,\end{equation}
where $\gamma_1$ and $\gamma_2$ are the components of the complex shear. The shear strength is $\gamma_{\text{ext}} = \sqrt{\gamma_1^2 + \gamma^2_2}$ and its orientation is $\phi_{\text{ext}} = \frac{1}{2}\arctan(\gamma_2 / \gamma_1)$. The first and second derivatives of the potential can easily be determined by analytical derivation of the lensing potential definition.

\subsection{Lens mass profile: In practice}
\label{sect_lens_mass_profile_practice}

\subsubsection{Strategy}

In practice, the lens mass profile consisting of elliptical slices needs to be defined by characterizing each slice. The choice of the azimuthal profile given to the slices depends on the scientific question that is to be answered. Beyond the general investigation of the impact of twists and ellipticity gradients on lensing analyses, we particularly investigated which types of isodensity perturbations, that is, twists or ellipticity gradients, affect the fitted mass density profile most. We determined whether the same effects are produced if the variations are located in different regions of the galaxy. We further examined whether the combination of both position angle and ellipticity variations created unexpected degeneracies in the model. The quantitative impacts of the different perturbations when realistic twists and ellipticity gradients were considered were also studied.

\begin{figure*}
    \centering
    \includegraphics[width=\textwidth]{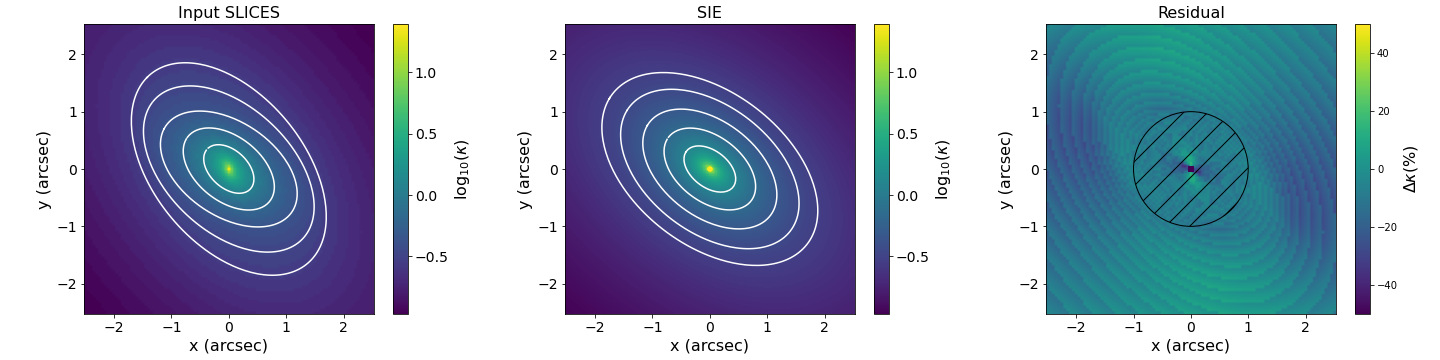}
    \includegraphics[width=\textwidth]{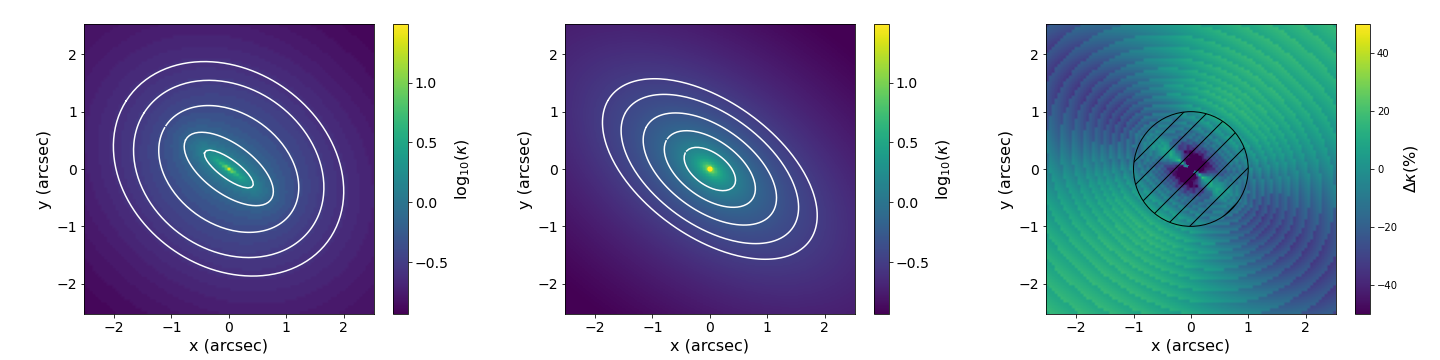}
    \caption{Comparison of mass profiles displaying twists or ellipticity variations with a mass profile following a pure ellipsoid. Left: Example of input profiles created with the slice method as explained in Sect. \ref{sect_lens_mass_prof_theory}, displaying either twists (top) or ellipticity changes (bottom). Middle: Singular isothermal ellipsoid with the ellipticity and position angle of the corresponding sliced profile at the Einstein radius, i.e., 1$\arcsec$. Right: Difference of the two convergence maps in percentage of the input profile. The dashed region indicates the region inside the Einstein radius.  }
    \label{fig_example}
\end{figure*}

We expect the twists to create quadrupolar moments that vary in orientation with radius; see the example in Fig. \ref{fig_example} (top). The twists outside the Einstein radius are thus expected to be equivalent to additional external shears, while those inside the Einstein radius are expected to mostly influence the orientation of the fitted ellipsoid. Variations in ellipticity are expected to change the strength of quadrupolar moment with radius, but not its orientation, as illustrated in Fig. \ref{fig_example} (bottom). The ellipticity gradient is consequently absorbed by an interplay of shear and ellipsoid azimuthal shape, and may influence the slope of the radial profile of the ellipsoid to better balance the internal versus external shear, as suggested by \cite{Kochanek2021}. 

To answer the different questions we raised, we explored two main paths: 1) We created an experiment in which we controlled the position angle profiles and the ellipticity gradients in order to systematically determine the influence of either twists or ellipticity changes in different parts of the galaxy, that is, inside the Einstein radius, at the Einstein radius, and outside the Einstein radius.  2) We created samples with twists and ellipticity changes based on realistic galaxies to investigate the impact of azimuthal structures on the population level. We note that by convention, when we report quantities measured at the Einstein radius, we effectively provide its value as measured at the slice whose major axis is closest to the Einstein radius.

Our first experiment consisted of using linear functions to describe the position angles or ellipticity profile, allowing changes in azimuthal structures at specific locations in the mock galaxy. This first sample was not meant to reproduce real galaxies, but limited the azimuthal structure to specific and controlled aspects. This allowed us to qualitatively assess the effect of these structures on the modeling behavior. 

The second experiment we conducted mimicked real galaxies. We used two physically motivated samples: We first used observation-based morphological properties of nearby elliptical galaxies, which has the drawback of only tracing the light. We then used morphological properties of hydrodynamically simulated galaxies, which allowed us to trace the full mass profile, but might not always match the mass profiles of real galaxies. With each sample, we used elliptical slices to create mock lensing-mass profiles for which we selected either twists or ellipticity gradients to analyze each effect separately. 
We then combined the two variations to mimic populations that encompass the diversity of a sample of real lensing galaxies.  

\subsubsection{Monotonic azimuthal variations}
\label{sect_lens_mass_prof_controlled_experiment}

First, we constructed a lens with slices displaying position angles and/or ellipticities that varied with radius according to an analytical function. The purpose of this first way of producing perturbations was to exactly control the input twist or ellipticities and acquire a basic knowledge about the influence of specific twist and ellipticity changes in lensing images.

\begin{figure}
    \centering
    \includegraphics[width=0.49\textwidth]{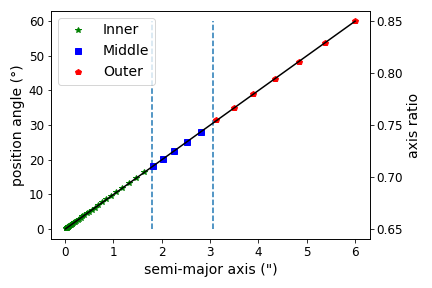}
     \caption{Slice properties of a fully varying profile in the monotonic azimuthal variations experiment, with either twists (left y-axis) or ellipticity gradient (right y-axis). The dashed vertical lines show from left to right the line at which the semi-major axis = 0.9~$\theta_E$ and the line at which the semi-minor axis = 1.1~$\theta_E$, considering an Einstein radius of 2\arcsec. They separate the different regions considered for variations (see Sect. \ref{sect_lens_mass_prof_controlled_experiment}). The slices belonging to the different regions are represented with different colors and symbols.}
    \label{fig_in_mid_out_profile}
\end{figure}

In this test, we assumed a linear change in position angles or ellipticity with semi-major axis. We considered both variations separately. We chose a profile made of 60 slices, evenly spaced in log space, ranging from 0.01$\arcsec$ to 6$\arcsec$. When we varied the ellipticity, the latter grew linearly from 0.65 to 0.85, that is, at a rate of 0.033 per arcsec. When instead twists were considered, the angles varied linearly from 0$\degree$ to 60 $\degree$, that is, 10$\degree$ per arcsec rate of change (see Fig. \ref{fig_in_mid_out_profile} and Fig. \ref{fig_example_profile}). This variation is not meant to be realistic, but is still comparable to what is observed in hydrosimulated galaxies or in local galaxies, as shown in Fig. \ref{fig_example_profile}. The Einstein radius was set to $\theta_E=2\arcsec$, such that the variations in ellipticity or positions angles were significant both inside and outside the Einstein radius (see Fig. \ref{fig_in_mid_out_profile}). 

We considered three regions in which variations could occur: the inner region, that is, the region in which the semi-major axis of the slices is < 0.9 $\theta_E$ ; the middle region, characterized by a semi-major axis > 0.9 $\theta_E$ and semi-minor axis < 1.1 $\theta_E$ ; and the outer region, in which the semi-minor axis > 1.1 $\theta_E$. With these regions, we created five different types of mock lensing profiles:
(1)  Mocks in which all slices were aligned and displayed the same ellipticity, that is, a constant axis-ratio and position angle in all regions. (2) Mocks for which only the inner part of the input galaxy was allowed to vary, either with twists or ellipticity changes. In other words, we varied the inner region, but the middle and outer regions remained constant.(3)  Mocks with changes only near the Einstein radius, that is, changes in the middle region and a constant profile in the inner and outer regions. (4) Mocks in which only the outskirts of the galaxy varied, that is, variations in the outer region, but not in the inner and middle regions. (5) Finally, mocks in which variations occurred in all the three regions.

When either the position angle or the ellipticity was not allowed to vary, the fixed value was chosen to be the one at the Einstein radius. Figure \ref{fig_in_mid_out_profile} shows the fully varying position angle or axis ratio profile of the slices as a function of semi-major axis. The three regions we considered to delimit the different cases described above are indicated.

\subsubsection{Data-motivated azimuthal variations}
\label{sect_realistic_samples}
Second, we used realistic changes of ellipticity and/or position angles to quantitatively measure the impact of these variations on the lens modeling parameters. We investigated two types of realistic lensing galaxy samples. (a) We used observed light profiles of local massive elliptical galaxies through the sample studied by \cite{Kormendy2009}. As the dark matter of these galaxies cannot be measured directly, the isophotes are the best reasonable proxy to the shape of the mass profile of elliptical galaxies. (b) We used hydrodynamical simulations of the universe and directly accessed the mass of massive ellipticals at redshifts typical of lensing galaxies with the EAGLE\footnote{EAGLE stands for Evolution and Assembly of GaLaxies and their Environments} simulations \citep{Schaye2015,Crain2015,McAlpine2016}. The two samples have differences and similarities, which highlights the necessity of exploring both. We first present the two samples and then compare them.

For each galaxy in the sample, we created three types of profiles: (1) allowing variation of ellipticity, but fixing the position angles, (2) allowing twists, but fixing the ellipticity, and (3) allowing both types of variations. Fixing either the positions angles or the ellipticity helps understanding the role of each type of variations for the lensing quantities through the modeling scheme, while the cases where both types of changes are displayed are most representative of massive elliptical galaxies. For comparison purposes, a fiducial profile with both fixed position angle and ellipticity was also created for each galaxy.
Since azimuthal structure variations may not be self-similar with galactocentric radii, we created mocks for systems with Einstein radii of either 1 or 2 $\arcsec$.
\paragraph{(a) Observation-based morphologies}
\label{sect_lens_mass_prof_observed_sample}
~\\\\

\cite{Kormendy2009} analyzed the light profile of all known early-type galaxies in the Virgo cluster and provided the isophotal analysis of each galaxy (i.e., size, ellipticity, and orientation of ellipses matching the isophotes constituting the light profile). Among others, they included lenticular galaxies, dwarf ellipticals, and spheroidal galaxies \citep{Kormendy2009,Graham2019}. We did not select any of these peculiar types of early-type galaxies. We selected nine massive elliptical galaxies that extend at least up to 4$\arcsec$ when redshifted to the target lens redshift (see Sect. \ref{sect_mock_images} for a discussion of the redshift). While this limit of 4$\arcsec$ may seem arbitrary, it ensures that the galaxies are wide enough to provide isophotal data up to at least twice the Einstein radius. Our sample is thus composed of the following elliptical galaxies: NGC 4649, NGC 4374, NGC 4261, NGC 4382, NGC 4636, NGC 4459, NGC 4473, NGC 4472, and NGC 4486. 

The advantage of this sample is that it represents light measurements of real galaxies. Even without knowing the dark matter distributions, we can at least be sure that this sample is representative of the light distributions of typical local massive elliptical galaxies.

\paragraph{(b) Hydro-simulation based morphologies}
\label{sect_lens_mass_prof_simulated_sample}
~\\\\

The EAGLE project \citep{Schaye2015,Crain2015,McAlpine2016} is a suite of hydrodynamic simulations that was run with a modified version of the smoothed particle hydrodynamics code \texttt{GADGET3} (last described by \citealt{springel2005b}) with $1504^3$ dark matter particles and an equal number of baryonic particles. The gravitational softening length of these particles is 2.66 comoving kpc (ckpc), limited to a maximum physical scale of 0.7 proper kpc (pkpc). The resulting galaxies are in overall agreement with observed properties such as the star formation rate, passive fraction, Tully-Fischer relation, total stellar luminosity of galaxy cluster and colors \citep{Schaye2015,trayford2015}, the evolution of the galaxy stellar mass function and sizes \citep{furlong2015a,furlong2017}, rotation curves \citep{schaller2015a}, and the $\alpha$-enhancement of early-type galaxies \citep{Segers2016}. 

To select a sample similarly populated as the \cite{Kormendy2009} sample, we used several criteria. First, we selected galaxies at redshift $z=0.271$ (see Sect. \ref{sect_mock_images} for a discussion of redshift), and then used a threshold at $10^{11} \text{M}_\odot$ to select only the most massive galaxies. To avoid most spiral or lenticular galaxies displaying disky shapes, we considered the galaxies properties and selected only those with a principal axial ratio $c/b>0.7$ following \cite{Trayford2019}. We then applied the Simulating EAGLE LEnses \citep[SEAGLE;][]{Mukherjee2018SEAGLEI} lens-simulation pipeline \citep{Mukherjee2018, Mukherjee2019}, which uses the \texttt{GLAMER} code \citep{GLAMERI,GLAMERII} for particles projection, to the selected galaxies to create their dark matter, stellar, and gas surface mass density maps using different projection axes. The sum of the three components is the total projected mass map of a galaxy. Our selection still comprised about 25 galaxies, with projected mass maps in three arbitrary orthogonal directions. We thus randomly chose a subsample of 12 mass maps out of these. Three of these have peculiar ellipticities in the center of the galaxy, that is, axis ratios lower than 0.3 at radii typically smaller than 0.8$\arcsec$. This intermediate-scale disk component is typical of ES, also called ellicular, galaxies \citep{Liller1966,Graham2019_classification,Graham2019}. The three ES galaxies appear as ES only in one projection. In the other projections, they appear as regular elliptical galaxies. 

To convert the EAGLE mass maps into a series of elliptical slices, we used the \texttt{AutoProf} \citep{AutoProf} software. \texttt{AutoProf} is a recent software providing pipelines that allow reducing, treating, and analyzing several images at the same time. We only used the part of the pipeline that fits ellipses at different radii of a galaxy and display, among others, the ellipticity, the position angle, and the size of each retrieved ellipse. This isophotal fitting algorithm is based on \cite{Jedrzejewski1987} with improvements in speed and accuracy through procedures adapted from machine-learning techniques. We cross-validated the resulting profiles with the more broadly used function \texttt{Ellipse} in \texttt{photutils} \citep{photutils101}. This last method is similarly based on \cite{Jedrzejewski1987}, but does not include the regularization schemes borrowed from machine learning used in \texttt{AutoProf}. As expected following \cite{AutoProf} for regular elliptical galaxies, the results from the two methods are similar; the \texttt{AutoProf} results are more stable.


\paragraph{(c) Comparison of properties of the two realistic samples}
~\\\\ 
The two galaxy samples from observations and from hydro-simulations have slightly different properties in terms of changes in position angles and changes in ellipticity. To compare them, we introduced four metrics: $\Delta PA_{IN}$ ($\Delta elli_{IN}$), which is the change in position angle (ellipticity) between the slice with a semi-major axis at 0.25$\arcsec$ and the one at the Einstein radius ; and $\Delta PA_{OUT}$ ($\Delta elli_{OUT}$), being the change of position angle (ellipticity) between the slice with major axis at the Einstein radius and the one at twice this radius. These metrics summarize variations in ellipticity or position angles, even if the variation may not always be monotonic over the considered range of radii. The innermost limit is chosen to be 0.25$\arcsec$ to ensure that the shape parameters at this minimum semi-major axis are robust enough to be used as representative of the inner parts of the galaxies. Indeed, since the EAGLE mass maps have a pixel size of 0.05$\arcsec$, and the fitted ellipses used to recover the shape parameters have to span enough pixels to adequately retrieve the local shape.

\begin{figure*}
    \centering
    \includegraphics[width=0.7\textwidth]{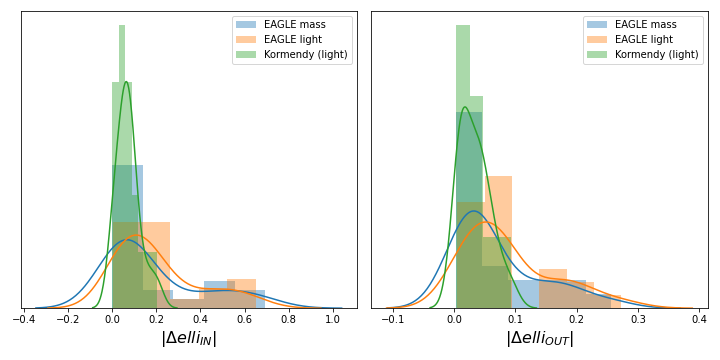}
    \includegraphics[width=0.7\textwidth]{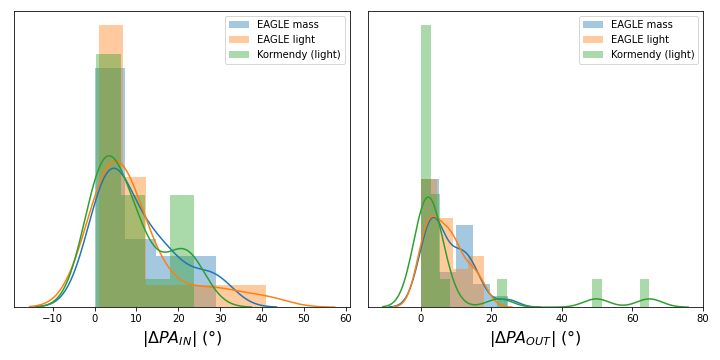}
    \caption{Distributions of absolute ellipticity changes (top) and position angles twists (bottom) for the observed and hydro-simulated sample. The left panels summarize changes between 0.25$\arcsec$ and the Einstein radius, and the right panels focus on changes between one and two Einstein radii. The observed sample, using the light of nearby galaxies analyzed by \cite{Kormendy2009} (green), is compared to the hydro-simulated sample using the mass of the EAGLE hydrodynamical simulations from \cite{Mukherjee2018} (blue). The light profiles associated with the simulated galaxies (orange) is also displayed to facilitate comparison between the two samples.}
    \label{fig_comp_korm_eagle_popu_twist_elli_changes}
\end{figure*}

Two galaxies in our sample extracted from \cite{Kormendy2009} display particularly drastic position angle twists. However, one of these two galaxies is almost circular with an axis ratio of 0.97 at $\theta_E$, compared to 0.84 for the other galaxy. Thus, only one galaxy in the chosen \cite{Kormendy2009} sample can be considered as significantly highly twisted. On the other hand, galaxies from the EAGLE sample generally display stronger changes in ellipticity. Since the sample from \cite{Kormendy2009} is based on isophotes of galaxies while the EAGLE sample is based on isodensity contours, the difference between the two sample could come from a difference between mass and light behavior. The EAGLE hydrodynamical simulations provide stellar mass maps in addition to total mass maps. 
We can thus analyze the light of the EAGLE galaxies and compare it to properties of the light of observed galaxies in a consistent way. Comparision between the different samples in terms of twists and ellipicity changes are shown at Fig. \ref{fig_comp_korm_eagle_popu_twist_elli_changes} as histograms and associated kernel density estimations (KDEs) of the four metrics for the samples we considered. The EAGLE isodensity and isophotal contours are similarly twisted and have analogous ellipticity changes. The range of ellipticity variations in the EAGLE sample, independently of whether we consider light or mass, is wider than the variation range from the sample from \cite{Kormendy2009}. However, the highest $\Delta elli_{IN}$ and $\Delta elli_{OUT}$ are due to the three ellicular galaxies that were probed at two different Einstein radius. Excluding these three galaxies would lead to still wider, but more similar distributions of the EAGLE and \cite{Kormendy2009} samples. We conclude that the differences between the two samples is mainly due to the different types of early-type galaxies that constitute them. For illustration purpose, we present at Fig. \ref{fig_example_profile} a set of position angle and axis ratio profiles for the linear experiment from Sect. \ref{sect_controlled_experiment}, two hydro-simulated galaxies, and two observed galaxies. 

\begin{figure}
    \centering
    \includegraphics[width=0.35\textwidth]{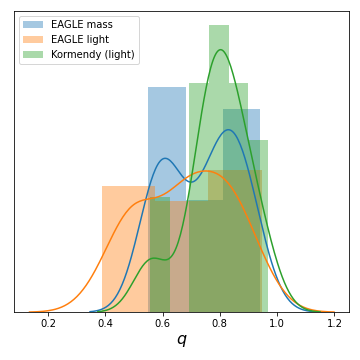}
    \caption{Distributions of the axis ratio at Einstein radius. The colors are the same as in Fig. \ref{fig_comp_korm_eagle_popu_twist_elli_changes}.}
    \label{fig_comp_elli_korm_eagle_popu}
\end{figure}

In addition to the position angle and ellipticity changes, we note that the axis ratios at the Einstein radius of the observed local galaxies are often higher than that of the simulated sample, as shown in Fig. \ref{fig_comp_elli_korm_eagle_popu}. This may yield a larger impact of the twists for the simulated sample, even with comparable distributions of the position angle variations.

\subsubsection{Radial mass distribution}
For each mock lensing galaxy, we adjusted the surface mass density of each slice such that the radial profile at the mean elliptical radii follows an isothermal profile. 
We also always fixed the center of every slice to be the same in order to avoid any lopsidedness and focus on the effect of changes of position angle and/or ellipticity with the galactocentric radius.

To better understand the interplay between the azimuthal changes and the radial mass profile, we also constructed mock images based on radial lensing profile following a composite mass profile with baryonic and dark matter components. The results of this experiment are presented in Appendix \ref{appendix_composite}.

Our method of specifying the constant surface density of each slice did not allow us to specify the Einstein radius and slopes with an arbitrary accuracy. This is particularly noticeable when there is a drastic change in ellipticity between several slices. In this case, the mass associated with the very elliptical slices in the inner part of the galaxy can be underestimated and create a shallower profile at radii typically smaller than a few pixel, which biases the overall Einstein radius. If these extreme variations also occur up to the Einstein radius, the slope at that radius can be affected. Only the ellicular galaxies are subject to these drastic ellipticity evolution. 
Generally, the simulated Einstein radius and the slope of the azimuthally averaged radial profile reach their targeted value (i.e., the value when no twist or ellipticity gradients are present) within an accuracy of 1\%. For the most extreme cases, the Einstein radius can be biased by up to 7\% and the slope by up to 3\%. Different schemes of mass attribution to the slices can slightly modify the strength of this effect, but cannot prevent it. When analyzing our results, we therefore ensured that we compared the fitted Einstein radius to the true input Einstein radius of each specific mock, calculated independently as being the radius at which the mean convergence drops below 1. This calculation has an accuracy of 0.01 \arcsec. The same practice was used for the slope results, which were compared to the input slope, measured using the azimuthally averaged logarithmic power-law slope of a profile at the Einstein radius. In addition to the mass profile of the lensing galaxy made of slices, a shear was added with a strength $\gamma_{\rm{ext}}=0.05$ and a random orientation $\phi_{\rm{ext}}$.

\subsection{Creation of mock images}
\label{sect_mock_images}

With the mass model produced as explained in the previous section, we used the \texttt{lenstronomy} software to create images of a background source composed of a quasar and a Sérsic host galaxy  ($R_{\rm sersic} =0.1, n_{\rm sersic}=3$) at redshift $z_s = 2$, lensed by the mock lensing galaxy. The source was randomly placed inside the inner caustic to create a quad lens in the image plane. The choice of a circular Sérsic source allowed us to avoid any specific degeneracy that could arise from the interplay between the ellipticity of the source, the image configuration, and the change in ellipticity or position angle in the lens. 

We always considered our lensing galaxy at redshift 0.271. This redshift is typical of lensing galaxies as observed in the SLACS survey \citep{Bolton2006,Treu2010,GonzalezNuevo2012} and was also used in the SEAGLE project \citep{Mukherjee2018SEAGLEI}, from which we used the pipeline to create mass maps in our sample of mock lensing galaxies based on hydro-dynamical simulations.

The lens was considered to be transparent to avoid blending between the lensed source images and the lens light. Incorporating lens light in our mock images would be more realistic. However, the lens light and the ring would overlap. Therefore the uncertainties introduced by the lens light subtraction would depend on the specific method used and would limit the generalization of the results. We nevertheless created a few mock images with lens light to assess the validity of our results for a nontransparent lens. We found that strong azimuthal variations in the lens light are generally not properly fitted by Sérsic profiles and ubiquitous patterns at the lens position can be seen in the residuals. On the other hand, if the lens light follows a nonazimuthally varying profile and is modeled as such, the additional freedom and noise introduced by the lens light beneath the arc do not influence the fitted parameters: consistent results are retrieved in both cases with and without lens light. We cannot rule out the possibility that specific lens-light profiles could degenerate with source-light patterns in the arc appearing due to azimuthal variations in the mass, and could bias the subsequent fit. However, such cases are expected to be rare.

We created mocks with typical space-based data quality. In particular, we chose to simulate observations obtained with the Wide Field Camera 3 (WFC3)  on board the Hubble Space Telescope (HST) in the F160W filter. This is one of the most frequently used setups for high-resolution imaging of lensed quasars \citep{Suyu2017H0licowI,Ding2021TDLMC}. We used a PSF created from the drizzling of eight PSFs extracted from real images. This PSF is the same as the one used in Rung 2 and 3 of the Time Delay Lens Modeling Challenge \citep[TDLMC;][]{Ding2021TDLMC}. We used a pixel size of 0.08$\arcsec$ and add noise assuming an exposure time of 5400\,s. Our setup is highly similar to other works that simulated lensing systems \citep[e.g., ][]{Ding2021TDLMC,Wagner2021,Park2021,multipole}. A noise map and a mask, masking the pixels with values lower than twice the background level, were also created and were used during the lens modeling procedure.

The time delay between the quasar images were also calculated within \texttt{lenstronomy}. We used the exact value of the delay in the modeling process, but associated a 2\% uncertainty with its value with a minimum threshold of one day.

\subsection{Fitting procedure}
\label{sect_fiiting_procedure}

After the mock images were created, they were modeled with \texttt{lenstronomy} assuming that the lensing galaxy is a singular power-law ellipsoid, and that an external shear is present. Since early-type galaxies density profiles are observed to be well approximated with a power-law profile up to large radii \citep{Suyu2009,suyu2010_B1608}, the power law was a standard model commonly employed to model real quasar images \citep{RXJ1131,Shajib_magnitudes,HolicowXIII}. 
We used a source model composed of a circular Sérsic profile with a quasar at its center.  
This setup uses similar light and mass profiles for the mocks and for the model, setting aside the azimuthal variations. It allowed us to test how the lack of flexibility of the model in the azimuthal direction impacted the inference. Because azimuthal and radial structures might interact, we later relaxed this assumption and used a radial distribution that no longer was a power law (Appendix \ref{appendix_composite}).


As a first step, the slope of the power-law ellipsoid mass distribution (PEMD) was fixed to simulate an isothermal profile. A particle swarm optimization \citep[PSO;][]{PSO1995,PSO1998} was then initiated to identify the maximum likelihood. After convergence was reached (or, in a few cases, maximum iteration reached), the slope constraint was relaxed and a second PSO was started. A Markov chain Monte Carlo (MCMC), through the software \texttt{emcee} \citep{Goodman2010emcee,Foreman-Mackey2013emcee}, finally proceeded in the vicinity of the second PSO optimum to sample the posterior and retrieve the uncertainties on the modeled parameters.

\section{Results}
\label{sect_results}




\subsection{Monotonic azimuthal variations}
\label{sect_controlled_experiment}


\begin{figure*}
    \centering
    \includegraphics[width=0.8\textwidth]{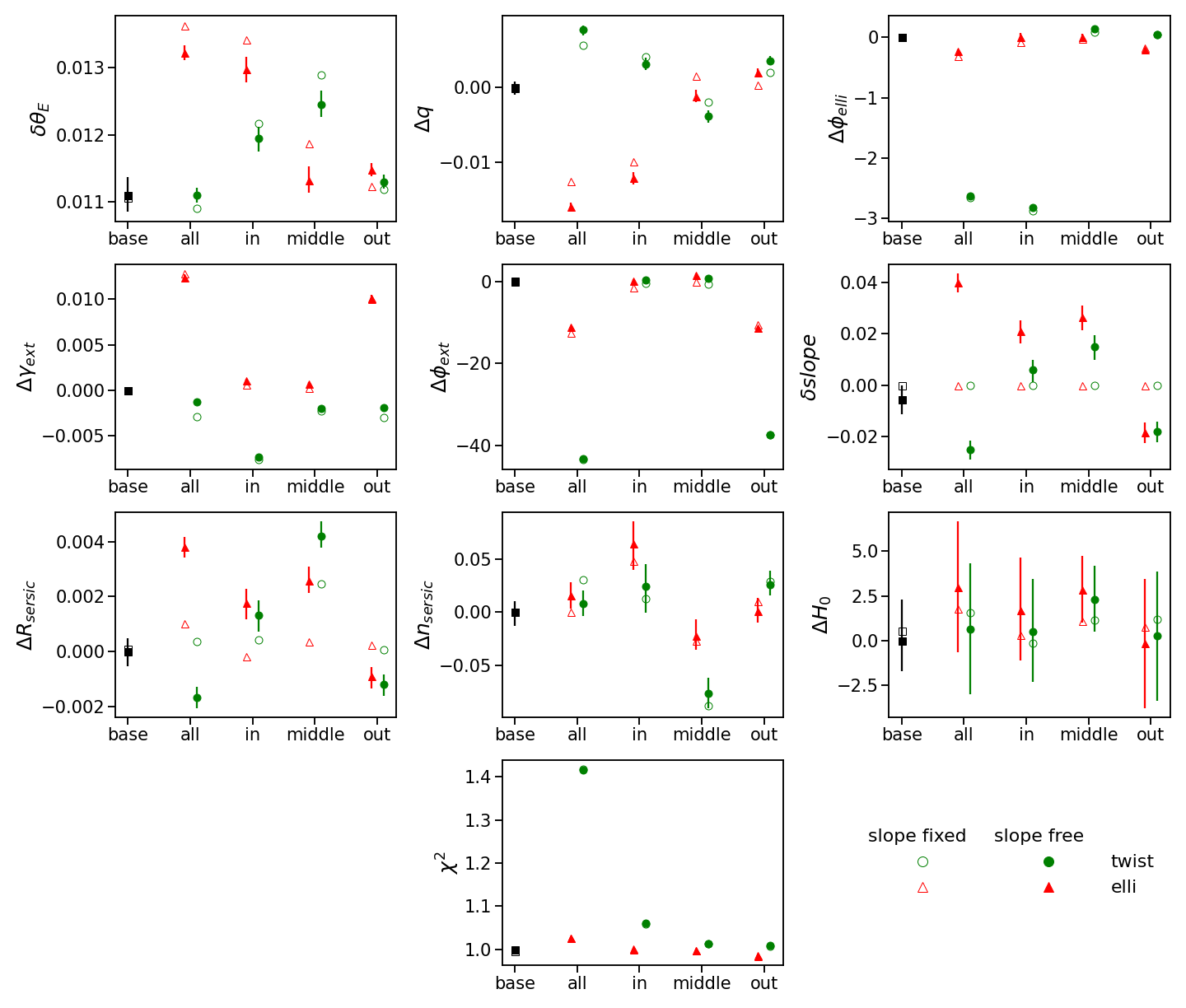}
    \caption{Fitted parameters of the performed experiment with linear variations of ellipticity and/or position angle with radius (see Sect. \ref{sect_controlled_experiment}). The two families of azimuthal variations, twist and ellipticity gradient, are displayed as green circles and red triangles, respectively. The results displayed with black squares are the ground-base case: the galaxy has isodensity contours that consist of ellipses displaying homoeidal symmetry (i.e., have the same axis ratio) and a single orientation. The open symbols characterize the results corresponding to the situation with a fixed fitted slope. The filled symbols display the median value of the final results after the constraint on the slope is relaxed. The error bars correspond to the 0.16 and 0.84 quantiles of the posterior distribution on the parameters. The x-axis tick labels indicate the type of variation considered: "base" refers to the ground-base case, "all" refers to variation occurring at all the radii, "in", "middle", and "out" indicate if the variation is limited to the region inside, around, or outside of the Einstein radius (see Sect. \ref{sect_lens_mass_prof_controlled_experiment} for exact definitions). The y-axis quantities are defined in Table \ref{table_description_param}.}
    \label{fig_in_out}
\end{figure*}

As described in Sect. \ref{sect_lens_mass_prof_controlled_experiment}, we performed an experiment with linear variations of ellipticity or position angles 
at radii lower, close to, and greater than the Einstein radius. While this test is not directly reflective of real galaxies, it is useful for understanding qualitatively the impact of each type of change occurring at different galactocentric radii of a lensing galaxy. The results of this experiment are displayed at Fig. \ref{fig_in_out}.

Figure~\ref{fig_in_out} shows that the changes in ellipticity outside the Einstein radius can be absorbed by the external shear by changing its strength by 0.01 and its orientation by 11\degree. The changes in ellipticity inside or at the Einstein radius, on the other hand, are absorbed by a change in slope that yields a bias of a few km/s/Mpc on $H_0$. 
The twists, on the other hand, mainly affect the retrieved orientations: the position angle of the ellipsoid is changed by up to 3 \degree for twists inside the Einstein radius, but the shear position angle varies by up to 37 \degree when the twist is outside $\theta_E$, in agreement with Fig. \ref{fig_example} (top). 
Twists present only at the Einstein radius do not influence orientations, as they are very localized and average out in the global mass profile. The effect of twists on the power-law slope and $H_0$ is harder to characterize as the evolution of both parameters when relaxing the fixed slope constraint is correlated, while the resulting $H_0$ remains mostly unbiased.

\subsection{Data-motivated azimuthal variations}
We now analyze the impact of twists and ellipticity gradients as they are displayed in data-motivated samples. The retrieved fitted parameters at a population level allow us to first verify the interpretation scheme outlined in Sect. \ref{sect_controlled_experiment}, and second, they quantify the effective impact of data-motivated azimuthal structures. This last outcome will inform us on any possible issue affecting lens models that ignore the lens azimuthal structure. We refer to Sect. \ref{sect_realistic_samples} for the description of the two samples used, that is, (a) a sample based on observed isophotes of local galaxies, and (b) a sample based on isodensity contours of galaxies drawn from hydrodynamical simulations.

\subsubsection{Observation-based morphologies}
\label{sect_korm_res}

\begin{figure*}
    \centering
    \includegraphics[width=\textwidth]{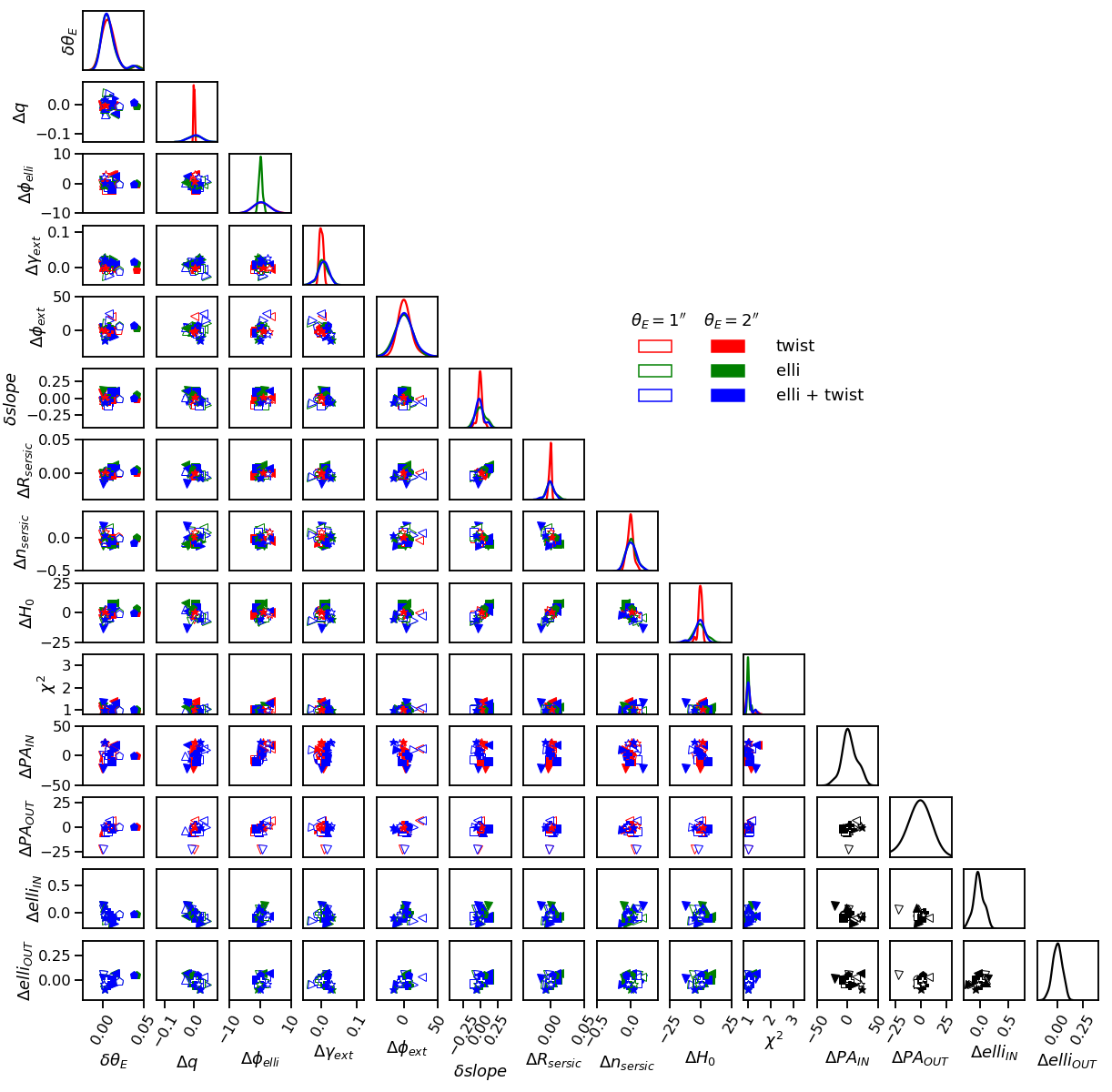}
    \caption{Population results based on mock mass profiles following the ellipticity and position angle variations observed in local galaxies, using the \cite{Kormendy2009} sample. The nine fiducial galaxies are represented with different symbols, filled when the considered Einstein radius is 2$\arcsec$, and empty for $\theta_E=1\arcsec$. Each galaxy is used to create three types of mock mass profile: the twist case in which only the position angles follow the original galaxy profile (red), the ellipticity case in which only changes of ellipticity are allowed (green), and the twist + ellipticity variation case (blue). The different mock images, created with the lens mass profile that lenses a circular Sérsic background source, are modeled with a PEMD + shear lens model and a circular Sérsic for the source model. The definition of the first nine quantities is listed in Table \ref{table_description_param}. The $\chi^2$ is the reduced imaging $\chi^2$ of the fit. The last four quantities are defined in Sect. \ref{sect_realistic_samples}. 
    Since these latter quantities are specific to the different galaxies and do not involve any modeling, they are plotted in black. 
    The diagonal cells are KDEs of the marginalized quantity. Only the fits with $\chi^2<1.5$ are considered for the KDE distributions.}
    \label{fig_korm_popu_res}
\end{figure*}

The summarized results of the selected \cite{Kormendy2009} population are displayed in Fig. \ref{fig_korm_popu_res} and Table \ref{tab_korm_res}. 
As described in Sect. \ref{sect_controlled_experiment}, the twists mainly impact the angles of the retrieved position angle of the main lens model, or the angle of the shear. The position angle of the ellipsoid differs from the fiducial angle with a scatter of $1.4\degree$. The position angle of the shear is offset with respect to the true shear angle with a scatter of $2.8\degree$ on average. We note that the amplitude of the position angle deviation for the shear may also depend on the strength and orientation of the fiducial input shear. The gradient of ellipticity, on the other hand, mainly impacts the shear strength, the slope, the source size, and the value of $H_0$, introducing a scatter of 22\%, 3.4\%, 5.0\%, and 5.1\%, respectively, on these quantities. We did not separate the $\theta_E=1\arcsec$ mock image results from the $\theta_E=2\arcsec$ results because the two subsamples are too small to distinguish a statistically significant difference of behavior.


Another clear trend is the correlation between $H_0$ and the fitted slope, but also with the source size. This suggests that to first order, the redistribution of the azimuthal mass due to ellipticity gradients has the same effect as the one produced by a mass-sheet transformation \citep[MST;][]{SS2013}, as the fitted model is degenerate with the data provided that the source is magnified. The fact that the $\chi^2$ is not 1 shows that this is not a perfect degeneracy. This behavior has been anticipated by \cite{Kochanek2021}, who stated that the presence of ellipticity variations will (1) favor a specific value of the fitted slope to balance internal and external shear introduced by the azimuthal perturbation and retrieve the correct general ring shape, (2) cause the model to increase or decrease the source size to recover the ring brightness and thickness, and (3) bias the value of $H_0$, which mainly depends on the lens mass model, accordingly. 

The dependence of the modeled ellipticity on the input changes of the ellipticity inside the Einstein radius is trivial. The slight correlation between position angle of the PEMD model and the amplitude of the input twist is also trivial. 
The distribution of $\chi^2$ informs us that good fits are retrieved in all cases, despite the twists and ellipticity changes. 
Finally, even if single $H_0$ values can be biased low or high, the recovered median value of the distribution is centered, within 0.3 $\sigma$, on the fiducial $H_0$, that is, 70 km/s/Mpc, for all types of azimuthal variations. We define $\sigma$ as the mean value of the two sides of the 68\% credible interval.

\begin{table}[]
    \centering
    \fontsize{9.}{12}\selectfont
    \begin{tabular}{c||c|c||c||l}
    Quantity & \multicolumn{3}{c||}{Difference between} & Quantity description\\
    &input&fiducial&fitted&\\\hline
    $\delta\theta_{\rm E}$&x&&x& Einstein radius [\arcsec]\\
    $\Delta q $&&x&x& Lens axis ratio\\
    $\Delta \phi_{\rm elli}$&&x&x& Lens position angle [\degree]\\
    $\Delta \gamma_{\rm ext}$&&x&x& Shear strength\\
    $\Delta \phi_{\rm ext}$&&x&x& Shear position angle [\degree]\\
    $\delta slope$&x&&x& Power-law slope\\
    $\Delta R_{\rm sersic}$&&x&x& Source Sérsic radius [\arcsec]\\
    $\Delta n_{\rm sersic}$&&x&x& Source Sérsic index\\
    \multirow{2}{*}{$\Delta H_0$}&&\multirow{2}{*}{x}&\multirow{2}{*}{x}&Hubble constant\\
    &&&&\multicolumn{1}{r}{[km/s/Mpc]}
    \end{tabular}
    \caption{Definition of the parameters used to compare models.}
    \tablefoot{"Fiducial" stands for the retrieved value of the parameter for the fiducial case without twists and no ellipticity changes. "Input" stands for the quantity calculated specifically for the mock, created with ellipticity changes and/or twits, taken at the Einstein radius for the effective slope.}
    \label{table_description_param}
\end{table}
\begin{table}[]
    \centering
    \fontsize{9.}{15}\selectfont
    \begin{tabular}{cccc}
    \hline \hline
    \multicolumn{4}{c}{Input}\\\hline
     & Twist & Elli & Elli + twist \\ \hline
    $\Delta PA_{IN}$ & $1.^{+13.}_{-6.}$ & - & $1.^{+13.}_{-6.}$\\\hline
    $\Delta PA_{OUT}$& $-0.6^{+2.2}_{-8.9}$ & - & $-0.6^{+2.2}_{-8.9}$\\\hline
    $\Delta elli_{IN}$& - & $-0.05^{+0.12}_{-0.04}$ & $-0.05^{+0.12}_{-0.04}$ \\\hline
    $\Delta elli_{OUT}$& - & $0.006^{+0.035}_{-0.050}$ & $0.006^{+0.035}_{-0.050}$\\
    \hline \hline
    \multicolumn{4}{c}{Results}\\
    \hline
     &  Twist & Elli & Elli + twist \\ \hline
$\delta \theta_E$  &  $ 6.5^{+ 5.8}_{ - 6.8} \times 10^{-3}$ &  $4.7^{+ 7.3}_{ - 3.6} \times 10^{-3} $ &  $ 5.8^{+ 7.1}_{ - 4.8} \times 10^{-3}$ \\ \hline 
$\Delta q $  &  $ 0.0^{+ 2.1}_{ - 2.3} \times 10^{-3}$ &  $0.0^{+ 2.1}_{ - 1.6} \times 10^{-2}$ &  $ 0.3^{+ 1.7}_{ - 1.8}\times 10^{-2}$ \\ \hline 
$\Delta\phi_{\rm elli}$  &  $ 0.1^{+ 1.6}_{ - 1.2} $ &  $1.7^{+ 3.4}_{ - 5.3} \times 10^{-1}$ &  $ 0.1^{+ 0.9}_{ - 1.0}$ \\ \hline 
$\Delta \gamma_{\rm ext}$  &  $ 0.0^{+ 5.1}_{ - 5.0} \times 10^{-3} $ &  $0.3^{+ 1.3}_{ - 0.8}\times 10^{-2} $ &  $ 0.7^{+ 0.9}_{ - 1.1}\times 10^{-2}$ \\ \hline 
$\Delta \phi_{\rm ext}$  &  $ -0.3^{+ 3.8}_{ - 1.9} $ &  $0.9^{+ 5.9}_{ - 4.8} $ &  $ 1.8^{+ 4.3}_{ - 9.1}$ \\ \hline 
$\delta slope$  &  $ -0.6^{+ 2.4}_{ - 1.6} \times 10^{-2} $ &  $-0.2^{+ 7.6}_{ - 6.1}\times 10^{-2} $ &  $ -1.4^{+ 3.4}_{ - 4.4}\times 10^{-2}$ \\ \hline
$\Delta R_{\rm sersic}$  &  $ -0.0^{+ 1.4}_{ - 1.8} \times 10^{-3}$ &  $-0.2^{+ 5.4}_{ - 4.6} \times 10^{-3} $ &  $ -0.8^{+ 4.5}_{ - 3.0}\times 10^{-3}$ \\ \hline
$\Delta n_{\rm sersic}$  &  $ -3.7^{+ 2.7}_{ - 4.3} \times 10^{-2}$ &  $0.2^{+ 6.1}_{ - 7.5}\times 10^{-2} $ &  $ 0.8^{+ 7.6}_{ - 6.3} \times 10^{-2}$ \\ \hline
$\Delta H_0$  &  $ -0.3^{+ 1.2}_{ - 0.9} $ &  $-0.8^{+ 4.0}_{ - 3.1} $ &  $ -0.6^{+ 2.2}_{ - 3.0}$ 
    \end{tabular}
    \caption{Kormendy population results}
    \tablefoot{Input sample properties (top) and statistics of the model results (bottom) for each azimuthal variation subsample. See Sect. \ref{sect_realistic_samples} for the comprehensive population description. See Table \ref{table_description_param} for the definition of the retrieved quantities and Fig. \ref{fig_korm_popu_res} for the visualization of the results. Only fits with $\chi^2<1.5$ are considered within the population results.}
    \label{tab_korm_res}
\end{table}

\subsubsection{Hydro-simulation based morphologies}
\label{sect_eagle_res}

\begin{figure*}
    \centering
    \includegraphics[width=\textwidth]{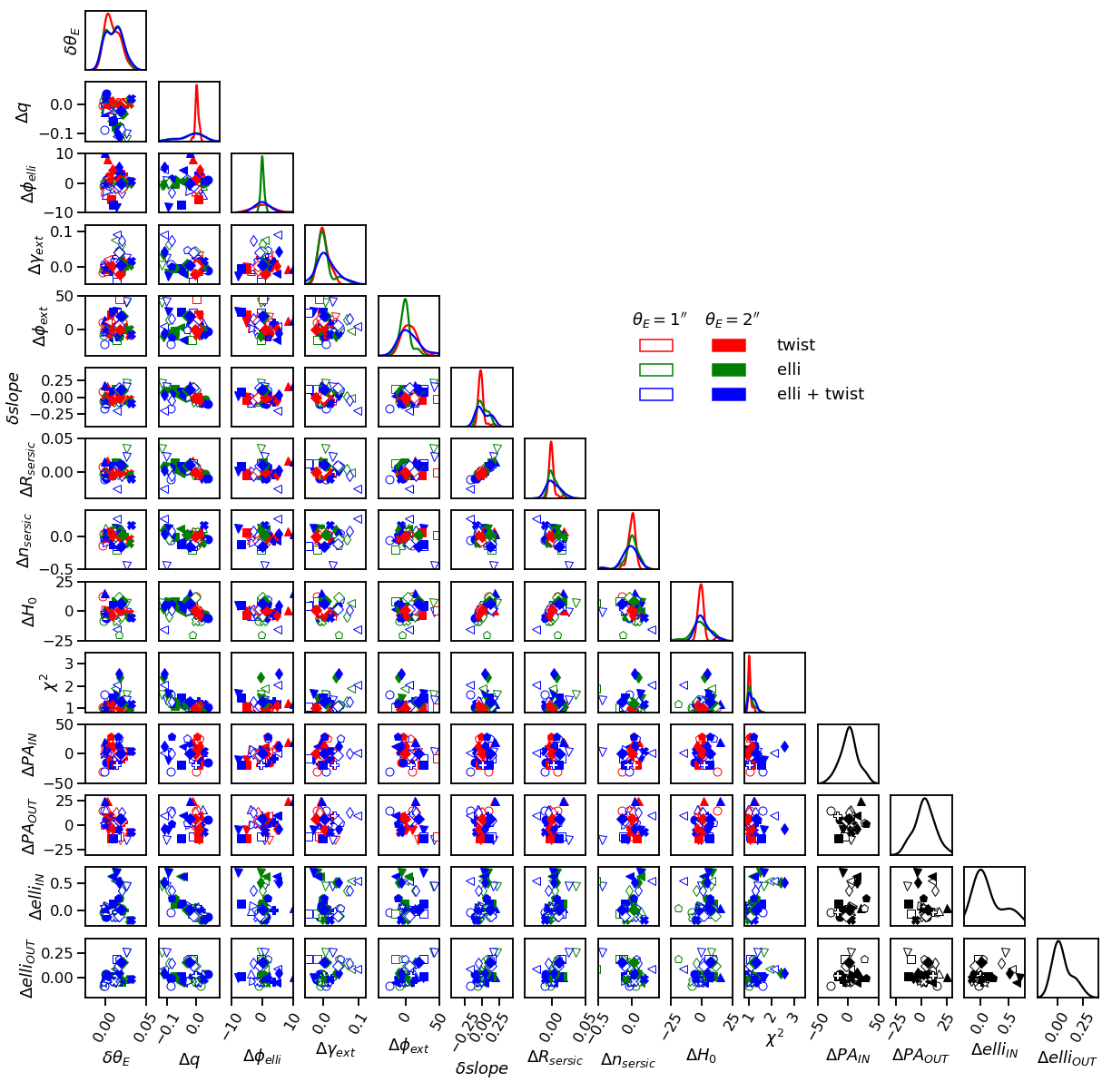}
    \caption{Population results for the EAGLE sample from \cite{Mukherjee2018}, based on 12 fiducial galaxies. The description is the same\textit{ \textup{as for Fig}. \ref{fig_korm_popu_res}.} The markers of the three ES galaxies are triangles pointing to the left, triangles pointing to the bottom, and the diamond shape. We recall that only the fits with $\chi^2<1.5$ are considered for the KDE distributions.}
    \label{fig_eagle_popu_res}
\end{figure*}

The results for the sample based on the EAGLE hydro-dynamical simulations is shown in Fig. \ref{fig_eagle_popu_res} and Table \ref{tab_eagle_res}. The observed trends are similar to those from the sample of \cite{Kormendy2009}, despite a few peculiarities that we explain below.

First, the scatter on the position angle of the ellipsoid and on the shear in presence of twists is wider than what we measured in Sect. \ref{sect_korm_res}, with population standard deviations of 2.4\degree \ and 8.4\degree , respectively. The scatter on the recovered shear strength is about 44\% in presence of ellipticity gradients, 26\% with twists, and reaches 50\% when the two morphological deformations are present, which is twice broader for this sample than for the sample of Kormendy. 
Second, when gradient of ellipticities are present, the retrieved axis ratio can differ by up to several percent from the ratio measured at the Einstein radius, in particular when a strong ellipticity gradient is present inside the Einstein ring. However, strong gradients are also harder to fit with a single elliptical profile, as indicated by the poorer $\chi^2$. The slope, $R_{\rm sersic}$, $n_{\rm sersic}$, and $H_0$ also have a broader distribution due to the ellipticity gradients than the twist-only galaxies: the parameter population distributions for ellipticity gradients are between 2.6 and 4.1 times broader than the twists gradients for the quantities cited above. 
Third, the correlation between the slope, $R_{\rm sersic}$, and $H_0$ is similar to the correlation reported earlier with the sample from \cite{Kormendy2009}. The correlation between the retrieved axis ratio and the changes in ellipticity inside the Einstein radius is trivial. 
Finally, the galaxies displaying the strongest ellipticity gradients, mostly the ellicular galaxies, are not well fit, as indicated by the $\chi^2$. The twist-only galaxies, on the other hand, are all fit well.

\begin{table}[]
    \centering
    \fontsize{9.}{15}\selectfont
        \begin{tabular}{cccc}
    \hline \hline
    \multicolumn{4}{c}{Input}\\\hline
    & Twist & Elli & Elli + twist \\ \hline
    $\Delta PA_{IN}$ & $2.^{+11.}_{-14.}$ & - & $2.^{+11.}_{-14.}$\\\hline
    $\Delta PA_{OUT}$ & $-2.8^{+8.5}_{-7.6}$ & - & $-2.8^{+8.5}_{-7.6}$\\\hline
    $\Delta elli_{IN}$ & - & $-0.04^{+0.42}_{-0.12}$ & $-0.04^{+0.42}_{-0.12}$ \\\hline
    $\Delta elli_{OUT}$ & - & $0.02^{+0.06}_{-0.13}$ & $0.02^{+0.06}_{-0.13}$\\
    \hline \hline
    \multicolumn{4}{c}{Results}\\
    \hline
    & Twist & Elli & Elli + twist \\ \hline
    
$\delta \theta_E$  &  $0.6^{+ 1.1}_{ - 0.5} \times 10^{-2} $ &  $1.2^{+ 0.8}_{ - 1.2} \times 10^{-2} $ &  $ 1.3^{+0.5}_{ - 1.3} \times 10^{-2}$ \\ \hline 
$\Delta q $  &  $ 1.5^{+ 6.8}_{ - 2.9} \times 10^{-3} $ &  $-1.2^{+ 2.7}_{ - 8.4} \times 10^{-2}$ &  $ -2.2^{+ 3.5}_{ - 6.8} \times 10^{-2}$ \\ \hline 
$\Delta\phi_{\rm elli}$  &  $ 0.4^{+ 2.3}_{ - 2.5} $ &  $-0.4^{+ 5.6}_{ - 4.3}  \times 10^{-1}$ &  $ -0.0^{+ 3.8}_{ - 3.2}$ \\ \hline 
$\Delta \gamma_{\rm ext}$  &  $ -0.0^{+ 1.6}_{ - 1.0} \times 10^{-2}$ &  $-0.1^{+ 4.0}_{ - 0.4} \times 10^{-2} $ &  $ 0.4^{+ 3.8}_{ - 1.2} \times 10^{-2}$ \\ \hline 
$\Delta \phi_{\rm ext}$  &  $ 7.6^{+ 7.0}_{ - 9.7} $ &  $-0.1^{+ 4.0}_{ - 6.9} $ &  $ 0.3^{+ 1.9}_{ - 0.9} \times 10^{1}$ \\ \hline 
$\delta slope$  &  $ -1.5^{+ 1.4}_{ - 2.6}\times 10^{-2} $ &  $0.6^{+ 9.9}_{ - 6.4} \times 10^{-2}$ &  $ -0.0^{+ 1.3}_{ - 0.7} \times 10^{-1}$ \\ \hline 
$\Delta R_{\rm sersic}$  &  $ -1.0^{+ 2.0}_{ - 1.7} \times 10^{-3} $ &  $0.9^{+ 7.2}_{ - 4.7} \times 10^{-3}$ &  $ 1.2^{+ 8.7}_{ - 7.9}\times 10^{-3}$ \\ \hline 
$\Delta n_{\rm sersic}$  &  $ 1.0^{+ 2.2}_{ - 5.1} \times 10^{-2}$ &  $0.0^{+ 0.8}_{ - 1.1} \times 10^{-1}$ &  $ -2.1^{+ 9.2}_{ - 8.7} \times 10^{-2}$ \\ \hline 
$\Delta H_0$  &  $ -0.5^{+ 1.3}_{ - 2.2} $ &  $0.0^{+ 8.1}_{ - 4.4} $ &  $ 1.3^{+ 5.2}_{ - 4.5}$    \end{tabular}
    \caption{EAGLE population results}
    \tablefoot{Distributions of azimuthal variations for the sample (top) and associated population results (bottom). See Sect. \ref{sect_realistic_samples} for a comprehensive description. See Table \ref{table_description_param} for the definition of the retrieved quantities and Fig. \ref{fig_eagle_popu_res} for the visualization of the results. Only fits with $\chi^2<1.5$ are considered.}
    \label{tab_eagle_res}
\end{table}

\subsubsection{Comparison of the two samples}
\begin{figure*}
    \centering
    \includegraphics[width=\textwidth]{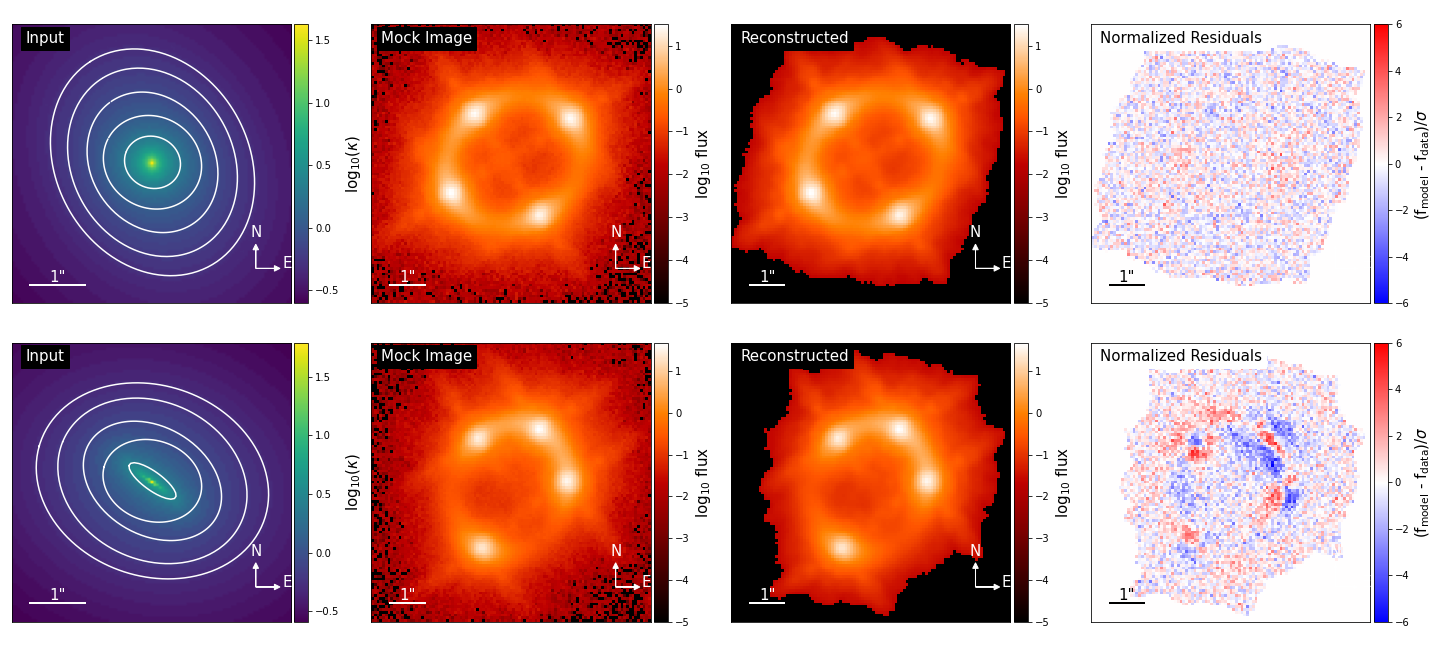}
    \caption{Example of a fit when ellipticity gradients and twists are well absorbed within the model (top) and when these azimuthal variations leave ubiquitous patterns in the residuals (bottom). From left to right: Convergence map of the input mass profile displaying twists and ellipticity gradients, mock image created with this mass profile, best power-law + shear model fitting the mock image, and residuals between the image and the model. In most cases, the residuals display no patterns. When extreme changes in ellipticity are displayed in the input mass model, as for ellicular galaxies, the azimuthal variations may not be absorbed by the model. In this case, the residuals display recognisable patterns that are unlikely to be mistaken as lens light or any other well-behaved light component.}
    \label{fig_fit_results_example}
\end{figure*}

\begin{figure*}
    \centering
    \includegraphics[width=\textwidth]{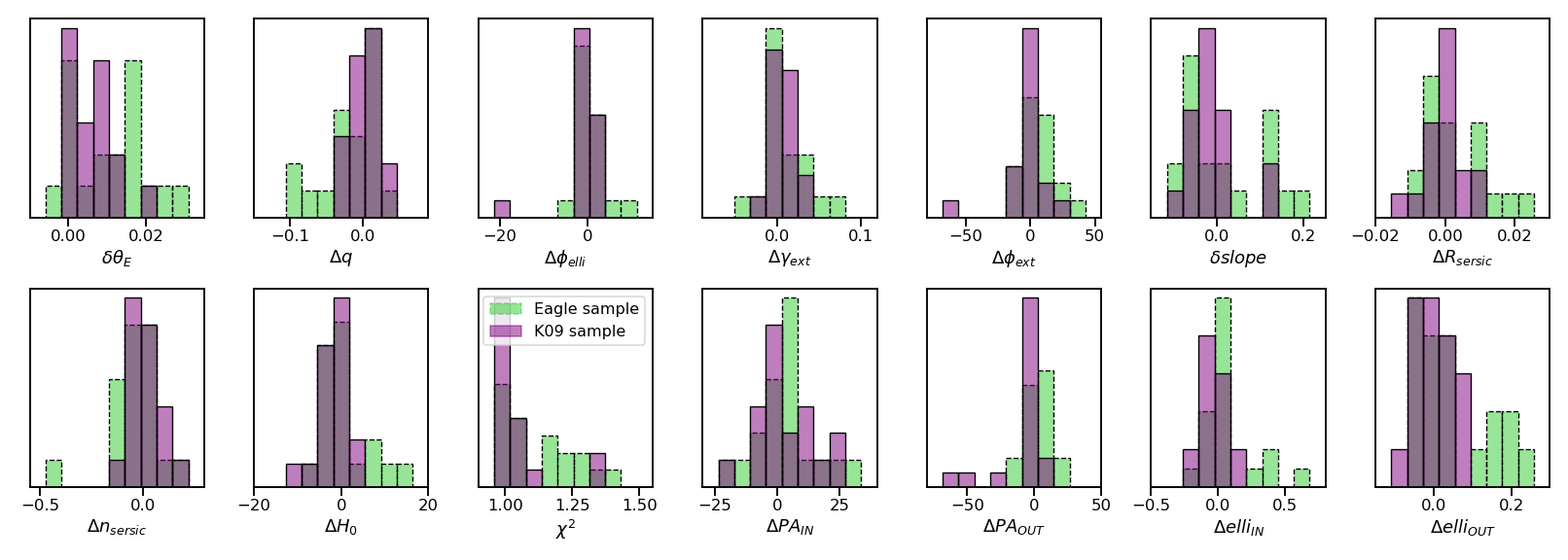}
    \caption{Comparison of results for the sample of mocks based on hydro-simulated morphologies (dashed green) and for the results based on observed morphologies (plain red) when both ellipticity variations and twists are present. Both variations are needed to mimic realistic galaxies. The displayed quantities are identical to those of Fig. \ref{fig_korm_popu_res}-\ref{fig_eagle_popu_res}. Only the fits displaying $\chi^2$<1.5 are considered.} 
    \label{fig_compa_eagle_korm_results}
\end{figure*}

We now compare the distributions of fitted quantities for our two mock populations in detail. Fig. \ref{fig_compa_eagle_korm_results} compares the histograms of the two population results when both twists and ellipticity changes are present. We anticipate that the combined set of EAGLE and Kormendy covers a wide range of azimuthal variations within which the real population is expected to lie. 
While the extreme changes in ellipticity are sometimes difficult to model, most twists and ellipticity changes are easily absorbed by the model, which is a simple power-law ellipsoid model and shear. Figure \ref{fig_fit_results_example} shows typical fit residuals when the azimuthal variations are well (top) and poorly (bottom) absorbed in the model. The galaxies based on the simulated sample display the most extreme ellipticity variations. The reason is that this sample includes ellicular galaxies, which are absent from the sample studied by Kormendy. For the two samples, the changes in position angles are absorbed by the position angle of the modeled ellipsoid and the orientation of the shear. The variations in ellipticity have an impact on the retrieved shear strength, with amplitudes depending on the sample. The same observation can be made for the slope of the power-law ellipsoid, the source size, and $H_0$. Moreover, no bias is observed, the recovered median value of $H_0$ being less than half a $\sigma$ away from the fiducial one. 
The $\Delta H_0$ distribution for the sample based on EAGLE galaxies is slightly broader, however, as they display a broader range of ellipticity variations (see Sect. \ref{sect_realistic_samples}).

The strong correlation between $H_0$, the power-law slope, and the source size when ellipticity gradients are present is similar to what would be observed in presence of  a mass sheet degeneracy (MSD), and it appears in both samples. This can be qualitatively understood from the bottom panel of Fig. \ref{fig_example}. To compensate for the missing or excess mass introduced by ellipticity gradients, the slope of the density profile of the model needs to be modified and rescaled to conserve the mass within the Einstein radius. A pure mass sheet transformation (MST) acts only as a rescaling of a \textit{\textup{radial}} mass density profile. Nevertheless, the ellipticity gradients, which are \textit{\textup{azimuthal}} variations, trigger an MST-like degeneracy. The pure MST effects have been mitigated so far: We used power-law mass profiles for both the input lens mass and the model, whereas an MST cannot transform a power law into a power law. It may therefore be important to relax the assumption on the radial density profile to better investigate the interplay between the radial mass profile and the azimuthal variations. We thus added Appendix \ref{appendix_composite} to explore this question further by creating mock lenses following a composite mass model instead of a power law. We did not observe an additional effect of the interaction of the radial and azimuthal mass profile. The MST due to the radial mass profile and the MST-like degeneracy due to azimuth variations seem to act independently.

\section{Summary and conclusions}
\label{sect_conclu}

Isophotal twists and ellipticity variations have been detected in elliptical galaxies in the local environment as well as at higher redshifts \citep[see, e.g.,][]{Fasano1989,Keeton2000Q0957,Hao2006,Pasquali2006,Kormendy2009}. Lensing galaxies, which mostly are massive ellipticals, can also display these features \citep[see, e.g.,][]{Keeton2000Q0957,Fadely2010Q0957}. However, most lensing galaxies are modeled with a simple ellipsoidal shape \citep[see, e.g.,][]{review_tdc,Shajib_magnitudes, HolicowXIII}. Characterizing the influence of this assumption has been the goal of this paper. 

We have followed a method similar to that of \cite{multipole}. 
We created mock lensing galaxies that displayed twists and/or ellipticity gradients using the method proposed by \cite{Schramm1994}. It consists of superposing finite elliptical slices of constant surface density, which have different sizes, mass, orientations, and ellipticities, to mimic an arbitrary mass profile. The lensing potential and deflection angle of each slice are known analytically, and were summed to compute the lensing quantities associated with the whole mass profile. We then simulated gravitationally lensed systems and modeled them with a mass profile characterized by a single orientation and ellipticity. The analysis of the retrieved parameters allowed us to quantify the impact of the scarcity of azimuthal structures in the lens modeling. 

We first simulated lensing galaxies with position angles and ellipticities that varied linearly with radius to acquire a heuristic understanding of the role of each of these two azimuthal structures on lensed images. We found that moderate azimuthal changes do not preclude good modeling of the systems. The twists in the inner regions of the galaxy mainly impact the orientation of the fitted ellipsoids, while twists in the outer regions of the galaxy rotate the position angle of the retrieved shear. For ellipticity gradients in the outer part of the galaxy, the shear strength and orientation are modified instead. Ellipticity gradients in the inner part mainly impact the ellipticity of the modeled density profile. Ellipticity variations also influence other parameters, however, such as the power-law slope, the source size, and the time-delay distance, which is a proxy for $H_0$. This is explained by a balance of the internal and external shear that is achieved in the modeling by an adjustment of the radial profile, thus modifying $H_0$, and accommodating the source size accordingly. This confirms the statement by \cite{Kochanek2021} that fitting a lensing galaxy displaying ellipticity gradients with a homoeidal power-law profile would lead to a bias on the slope to balance the internal and external shear introduced by the change of ellipticity, and would thus bias $H_0$. 

To quantify the impact of twists and gradients on lens parameters and cosmography, we simulated and modeled populations of lensing galaxies displaying realistic morphologies. Two populations of data-motivated azimuthal variations were considered: the light profiles of observed nearby galaxies \citep{Kormendy2009}, and the mass profiles of galaxies from the EAGLE hydrodynamical simulations \citep{Schaye2015}. Both samples display similar position angle variations. Ellicular galaxies in the EAGLE sample yield a wider population of ellipticity gradients, however, as the sample contains galaxies displaying drastic variations in their ellipticity profile (e.g., an axis ratio varying by 0.5 within 4~kpc). 
While extreme ellipticity variations are difficult to model, all the twists we tested were easily absorbed in the modeling. 
When both twists and ellipticity gradients are present in the mock lensing galaxy, the effect of each simply add up. The modeling of a single lens 
can reach a bias on the shear orientation of $20\degree$, while the shear strength can rise up to twice its fiducial value. The $H_0$ inference can also be biased up to 10 $\text{km}\,\text{s}^{-1}\,\text{Mpc}^{-1}$. Nevertheless, the influence of the azimuthal variations averages out at a population level. The median $H_0$ value is centered on the fiducial value within less than 0.5$\,\sigma$ for both observed and hydro-simulated morphologies. Other parameters such as the shear strength, shear orientation, and the ellipsoid axis ratio and position angle are also centered on their fiducial values at a population level. The absence of ellicular galaxies yields narrower distributions of the fitted parameters in the observation-based sample compared to the hydro-simulation sample. 


The reported results on a lens-by-lens case may cause concern, but fortunately, galaxies are not transparent, and twists and ellipticity gradients may be readily detectable in their luminosity profile. The lens-light variations are a sufficiently good proxy of the mass distribution azimuthal patterns according to the analysis we performed of elliptical galaxies from hydrodynamical simulations. By comparing light and mass in EAGLE, we found that the variations in orientation and axis ratio in the light are similar to the variation in the mass. The light is thus a good tracer of the position angle and ellipticity fluctuations in the inner 15 kpc of elliptical galaxies. The precision of the instrument point spread function, which blurs and circularizes the features of the galaxies, will nevertheless play a role in the detection of azimuthal variations, especially in the inner parts of the galaxy.

In principle, kinematic information should also provide additional constraining power that can help to inform and model the existence of such structures. However, in practice, neither the data quality nor the kinematic models commonly employed in lensing contexts are sufficient to this degree as yet. In the context of lens modeling, many assume spherical Jeans velocity dispersions \citep[e.g.,][]{HolicowIV,HolicowIX,HolicowXII}, or the most advanced models assume axisymmetry \citep[e.g.,][]{TDCOSMOVIII}, and therefore do not take azimuthal structures into account. While methods exist to create mock kinematics with additional structure (e.g., triaxiality via Schwarzschild modeling), these structures can currently not be modeled jointly with lens modeling. As kinematic simulation and lens modeling codes coevolve, it might be possible in the future to better constrain a complex galaxy with azimuthal variations using spatially resolved kinematic information.

To conclude, this paper explores the effect of isodensity twists and ellipticity gradients on lensing cosmography analyses, as such azimuthal variations exist in early-type galaxies. The impact of twists on the modeling is in general rather marginal. Ellipticity gradients, on the other hand, can introduce substantial amount of shear that can be partly absorbed by the lens models, with an impact on the value of $H_0$ on a single lens basis. However, no bias is observed at the population level. We did not construct our samples to perfectly match the true population of lensing galaxies, but we chose it with a rather broad diversity representative of real systems. An overpopulation of some subsamples might change the broad picture slightly, but the latter indicates that the impact of ellipticity gradients and twists is generally small or averages out. If a population analysis is not feasible, a lens-light analysis should enable a rough estimation of the azimuthal variations. It may be important to look at the light profile outside\textit{} the Einstein radius as ellipticity gradients in these regions can introduce substantial shear. If azimuthal variations are detectable in the isophotes, a more complex modeling scheme can consequently be applied to increase the azimuthal freedom in the mass model and assess the presence of a possible systematic bias.

More broadly, ellipticity gradients and twists are not the only type azimuthal variations that can be considered. \cite{multipole} also explored the role of undetected azimuthal variations in the mass on the model and specifically $H_0$, but focused on the role of the multipolar component of order 4, which corresponds to boxyness and diskyness. They found that boxyness or diskyness in the mass introduces specific patterns in the lensed arcs. Depending on the signal-to-noise ratio and on the complexity allowed in the source reconstruction, these patterns could be absorbed in the modeling with purely ellipsoidal power-law mass profiles. For ellipticity gradients and twists, the patterns are almost always unnoticed as they are easily modeled by changing the azimuthal components (orientation and strength of shear or  ellipticity) and/or by modifying the radial power-law mass model. Both ellipticity gradients and multipoles of order 4 have impacts on $H_0$ up to a bias of 15\%. However, a realistic combined population of boxy and disky galaxies averages the $H_0$ bias to be below the percent level, and similarly, a sample of early-type galaxies displaying a variety of ellipticity gradients also averages out the bias down to the percent level. A combination of all the different types of azimuthal variations (i.e., boxyness or diskyness, ellipticity gradients, and twists) was not explored in this paper. Nevertheless, as both multipolar variations and ellipticity gradients have little bias on $H_0$ at a population level, we expect that a sample of galaxies displaying a combination of both would also lead to an unbiased cosmological inference, even if cosmological results of individual lensing systems could be more widely spread.

\begin{acknowledgement}

The authors thank Frederic Courbin and Dandan Xu for the fruitful discussions on this work.

This work makes use of lenstronomy v1.9.1 \citep{lenstro2015,lenstro2018,lenstro2021}, AutoProf \citep{AutoProf}, and of the following Python packages : Python v3.6.5 \citep{Python1,Python2}, Astropy \citep{astropy:2013,astropy:2018}, Numpy \citep{Numpy}, Scipy \citep{scipy}, Matplotlib \citep{Matplotlib}, Pandas \citep{pandas1,pandas2}, Photutils \citep{photutils101}.

This project has received funding from the European Research
Council (ERC) under the European Union’s Horizon
2020 research and innovation programme (COSMICLENS :
grant agreement No 787886)
\end{acknowledgement}

\bibliographystyle{aa} 
\bibliography{biblio}

\appendix

\section{Example profiles}
\label{appendix_example_profiles}

\begin{figure}
    \centering
    \includegraphics[width=0.49\textwidth]{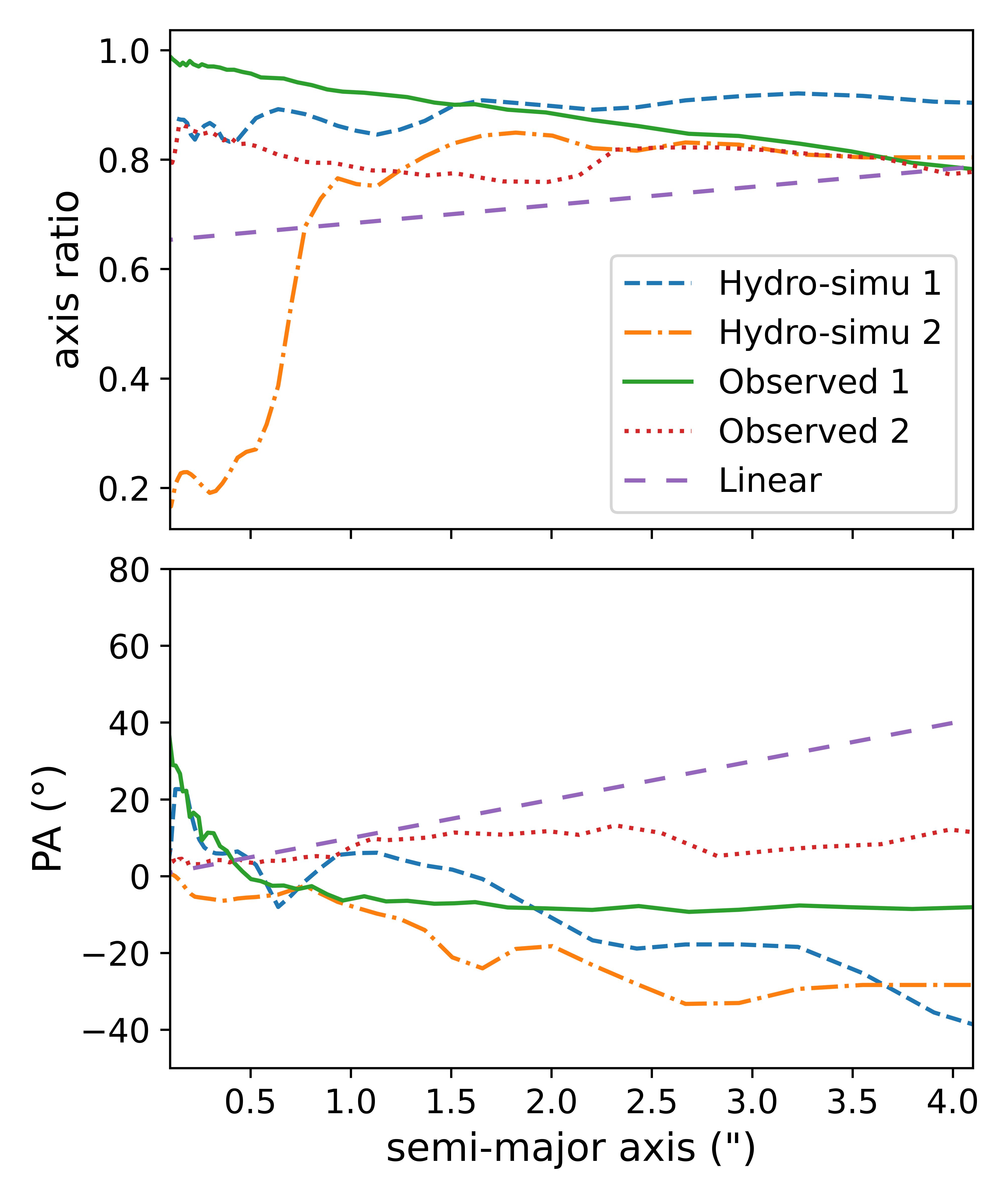}
    \caption{Example variations with semi-major axis of axis ratio (top) and position angle (bottom) from hydro-simulated, observed, and custom-designed galaxies. The "Hydro-simu 2" curves depict the morphology of an ellicular galaxy, i.e., an early-type galaxy displaying a medium-sized disk component in the center. The position angle profiles are translated by a constant angle to help visualization. The corresponding convergence maps are also shown in Fig. \ref{fig_example_kappa}.}
    \label{fig_example_profile}
\end{figure}

\begin{figure}
    \centering
    \includegraphics[width=0.49\textwidth]{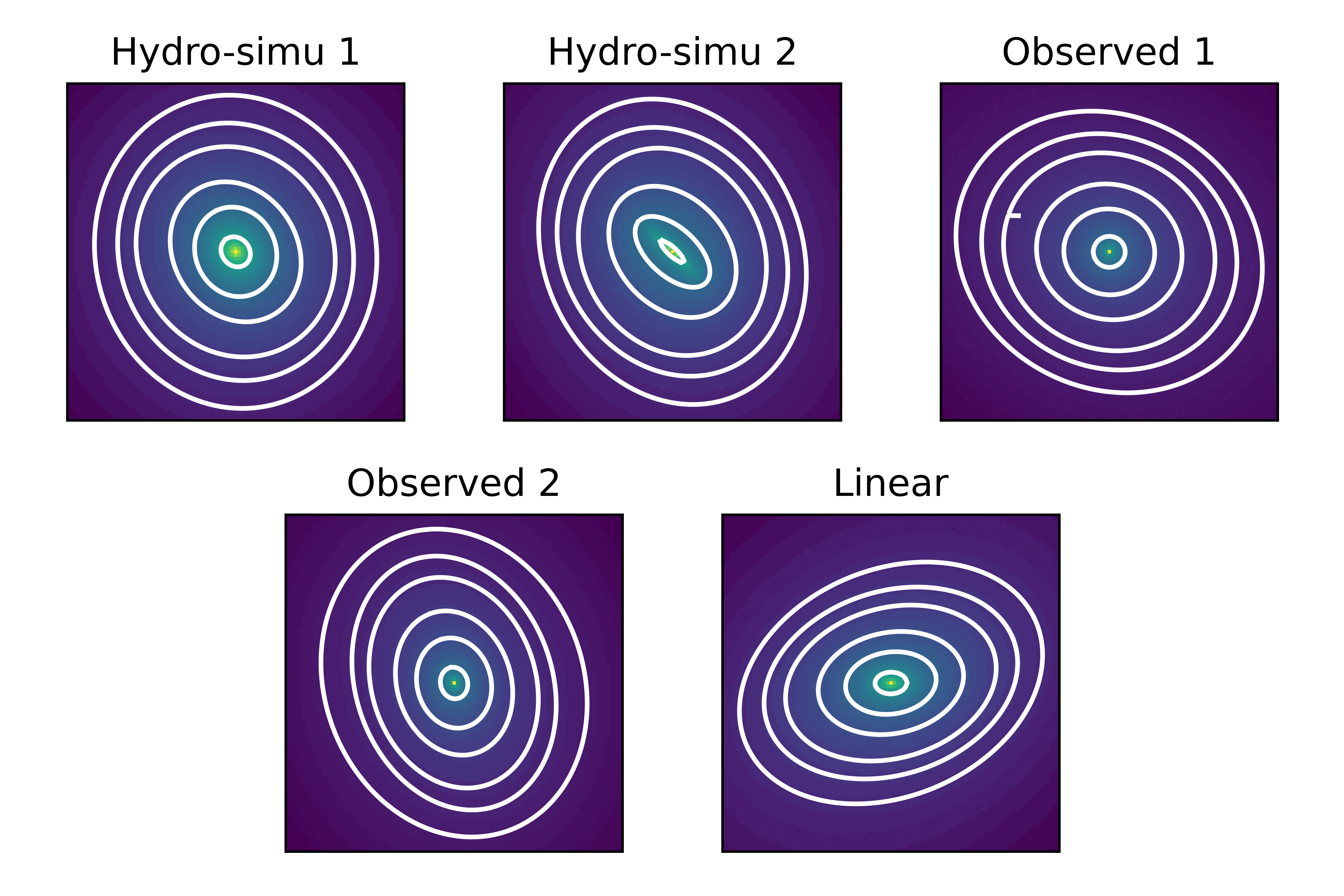}
    \caption{Examples of convergence maps created with slices for hydro-simulated, observed, and custom-designed galaxies that display axis ratio and position angle variations with semi-major axis. The corresponding azimuthal profiles of the different galaxies are displayed in Fig. \ref{fig_example_profile}.}
    \label{fig_example_kappa}
\end{figure}

We constructed mock lensing galaxies with varying azimuthal profiles. Five of these position angle and axis ratio profiles are presented at Fig. \ref{fig_example_profile}. The convergence maps created with slices following the different azimuthal variation profiles are shown in Fig. \ref{fig_example_kappa}. 
In Sect. \ref{sect_lens_mass_prof_controlled_experiment}, monotonic variations of position angle or ellipticity were considered. The profile used is displayed as a purple long-dashed line. Two data-motivated samples of azimuthal variations were considered in Sect. \ref{sect_realistic_samples}: the morphologies based on observed local elliptical galaxies analyzed by \cite{Kormendy2009}, and the morphologies based on the density of EAGLE hydro-simulated galaxies. Two profiles of each sample are presented in Fig. \ref{fig_example_profile}: with dashed blue and dash-dotted orange lines for the morphologies based on hydro-simulations, and as plain green or dotted red lines for those based on observations.


\section{Composite mass profiles}
\label{appendix_composite}
To explore the role of radial mass profile in conjunction with azimuthal variations such as twists or ellipticity gradients, we created new samples of mock lensing images based on composite lensing mass profiles consisting of a baryonic and a dark matter component. Unlike the power-law profile, this mass profile does not display  the same radial slope at every galactocentric radii. 

The baryonic component is usually represented with a Sérsic or a Chameleon profile \citep[][Gomer et al. in prep]{Dutton2011,Suyu2014}. The chameleon profile, which is the difference between two nonsingular isothermal ellipsoids (NIE), mimics the well-known Sérsic profile for Sérsic indexes between 1 and 4 and has the advantage of being less computationally demanding for the calculation of lensing quantities. The mathematical formulation we used for this profile is the following:
\begin{align}
    \kappa(\theta_1,\theta_2) = \frac{L_0}{1+q} \left(\frac{1}{\sqrt{\theta_1^2+\theta_2^2/q^2 + 4 w_c^2 /(1+q)^2 }}\right. -  &  \nonumber \\
    \left. \frac{1}{\sqrt{\theta_1^2+\theta_2^2/q^2 + 4 w_t^2 /(1+q)^2 }} \right),&
\end{align}
where ($\theta_1,\theta_2$) are defined within the coordinate system oriented along the main axis of the chameleon ellipsoid, $L_0$ is a normalization factor, $q$ is the axis ratio, $w_c$ is a proxy for the core radius of the first NIE, and $w_t$ is that of the second NIE. 

The dark matter component is commonly simulated with a Navaro-Frenk-White mass model \citep[NFW;][]{NavarroFrenkWhite1996}. The halo of dark matter thus follows
\begin{equation}
    \rho(r)=\frac{\rho_0}{(r/r_s)(1+r/r_s)^2}
,\end{equation}
where $\rho$ is the 3D mass density, $\rho_0$ the central mass density, and $r_s$ is the scale radius at which the profile smoothly switches between a $\rho \propto 1/r$ relation and a $\rho \propto 1/r^3$ proportionality. The lensing quantities formulation in 2D of this mass density in pseudo-elliptical symmetry can be found in \cite{GolseKneib2002}.

\begin{figure}
    \centering
    \includegraphics[width=0.49\textwidth]{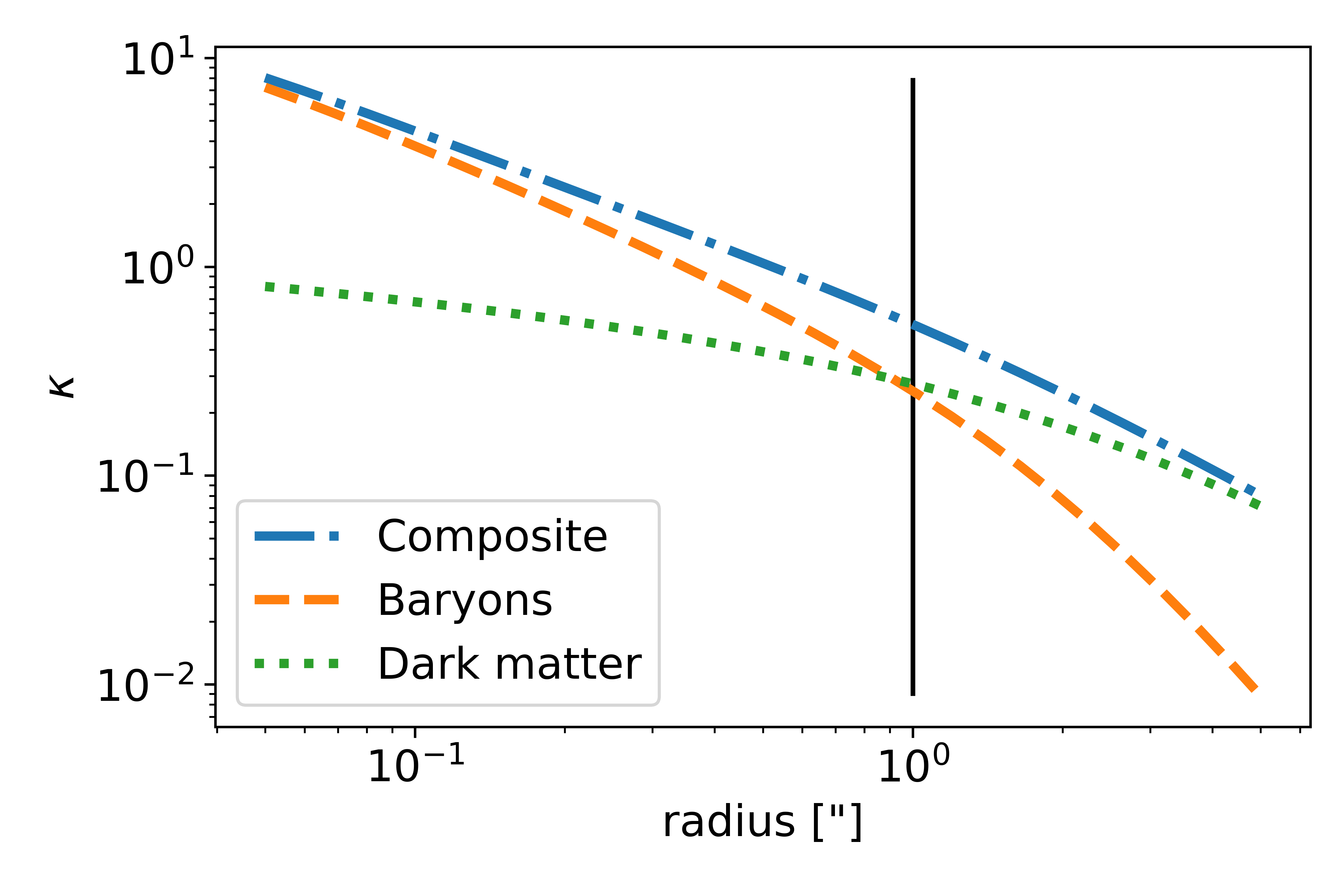}
    \caption{Radial profile of the composite mass model (dash-dotted blue line), made of a baryonic component following a chameleon profile (dashed orange line) and a dark matter component simulated with an NFW mass profile (dotted green line). The vertical black line indicates the Einstein radius.}
    \label{fig_composite_radial}
\end{figure}

The composite mass profile we chose to simulate had the following characteristics. The Einstein radius was set to 1$\arcsec$, which is typical of observed Einstein radii. Other characteristics were chosen such that in the end, the baryonic profile mimicked a Sérsic with $R_{\rm eff}=1.7 \arcsec$ and $n_{\rm sersic}=4$, corresponding to a total mass of $0.28\times 10^{12}~\textup{M}_\odot$, and the dark matter profile had physical size of $R_{200}= 0.25~\textup{Mpc}$ and $M_{200} = 2.4\times 10^{12}~\textup{M}_\odot $. The latter is defined as the mass enclosed in the radius $R_{200}$ in which the average density is 200 times the critical density of the universe at the considered redshift. This setup is arbitrary but representative of real lensing galaxies (Gomer et al. in prep). Such a radial mass profile is presented in Fig. \ref{fig_composite_radial}. 

\begin{figure}
    \centering
    \includegraphics[width=0.49\textwidth]{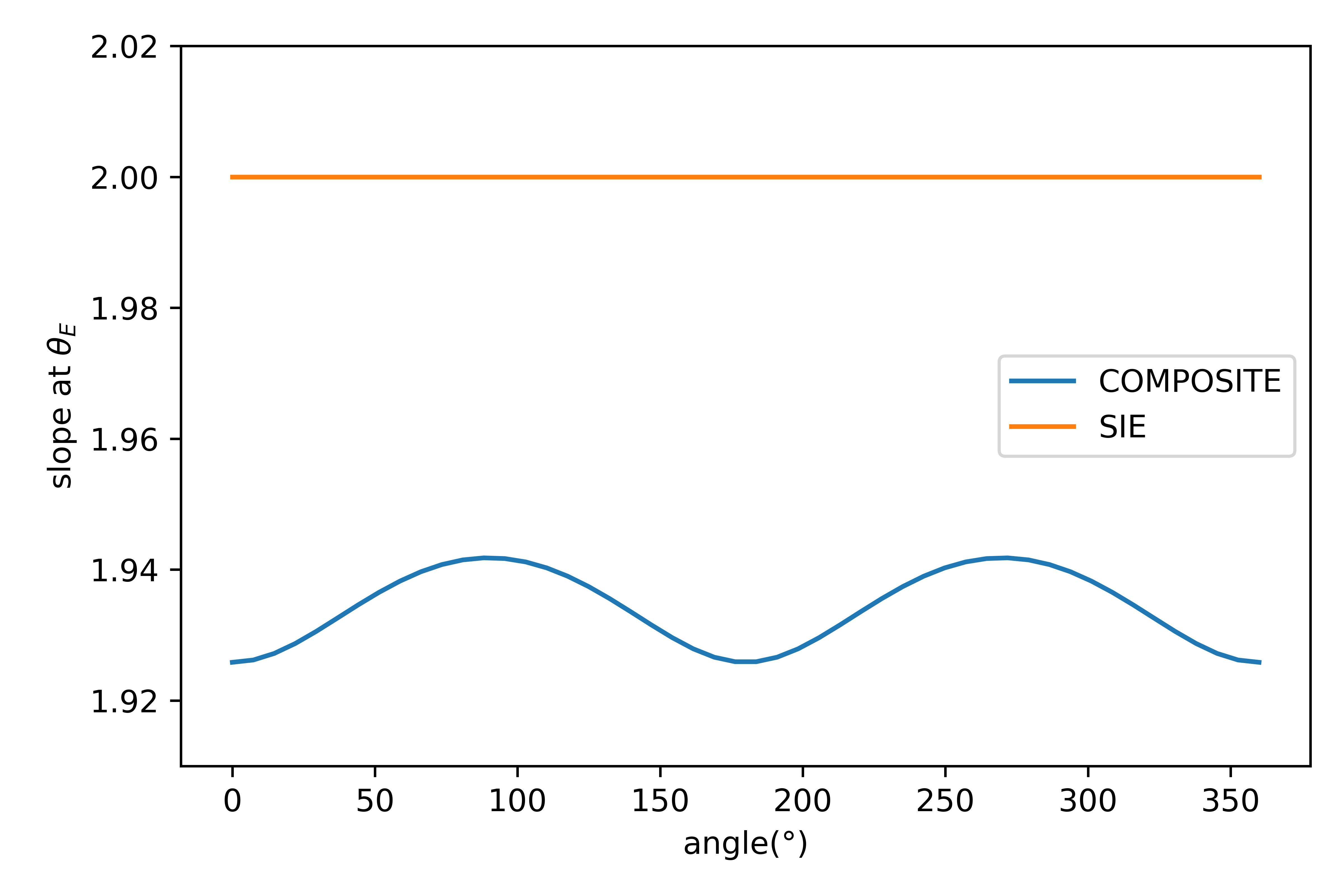}
        \caption{Circular azimuth cut of the profile slope for an SIE (orange) and the composite mass profile (blue) described in this section, considering a typical mass density axis ratio of 0.8 for both components, taken at the Einstein radius $\theta_E$=1$\arcsec$.}
    \label{fig_composite_azimuthal_slope}
\end{figure}

The radial mass profile was created using the azimuthal average of the elliptically symmetric mass density at each radius. However, the azimuthal profile shows that for an elliptical composite mass profile, the logarithmic slope of the profile is not constant for a circular azimuthal cut because it probes different elliptical radii of a profile with a varying slope with radius, as shown in Fig. \ref{fig_composite_azimuthal_slope}.

With the radial profile as simulated above, mock lensing images can be created for the different twist and ellipticity profiles of the two populations of data-motivated azimuthal variations, that is, the observation-based an the hydro-simulation-based morphologies. These images were then modeled with a PEMD and a shear, following the same procedure as implemented before.

\begin{figure*}
    \centering
    \includegraphics[width=\textwidth]{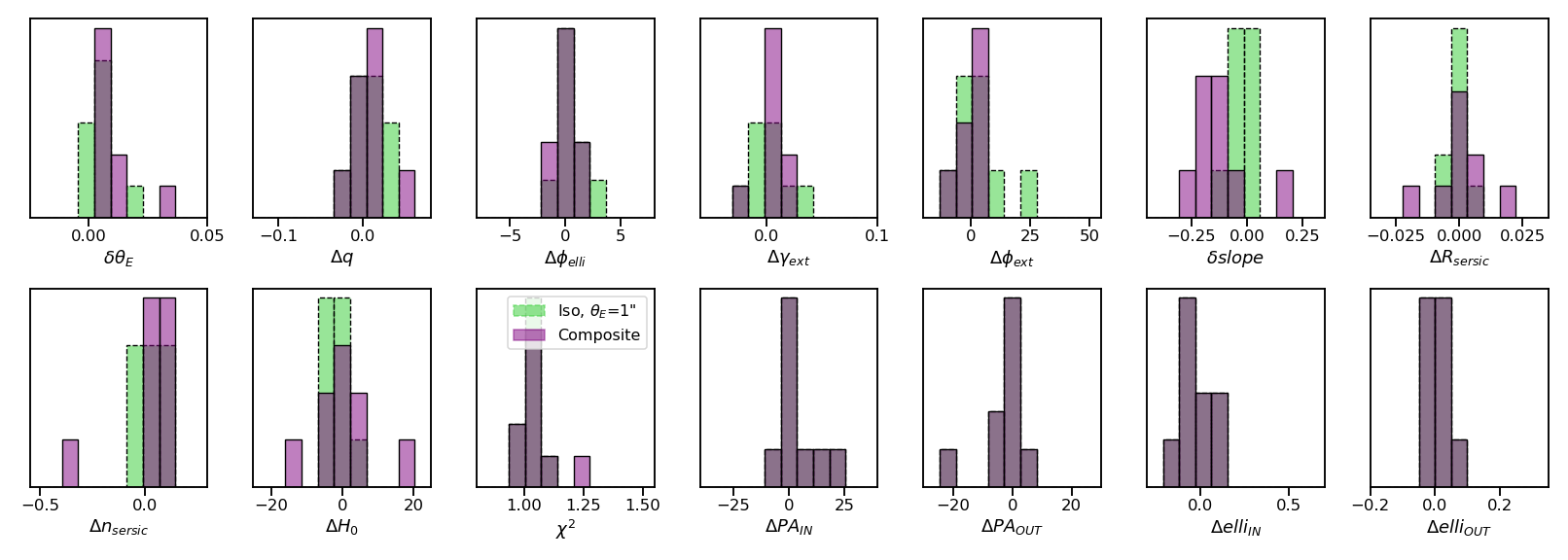}
    \caption{Comparison between fitted parameters of mock images created with either isothermal (dashed green) or composite radial mass profiles (plain red). The mock lensing mass models display both ellipticity changes and twists based on the \textbf{observed morphologies}, using the \cite{Kormendy2009} sample. Only $\theta_E=1\arcsec$ lensing galaxies are displayed for the isothermal cases in order to match the Einstein radius of the composite mock images samples. Only fits with resulting $\chi^2<1.5$ are considered.  The four right cells of the second row display identical distributions, as the same population of twists and ellipticity changes are used to create the composite and the SIE mocks, and all the fit results have sufficiently low $\chi^2$.}
    \label{fig_composite_korm_results}
\end{figure*}

\begin{figure*}
    \centering
    \includegraphics[width=\textwidth]{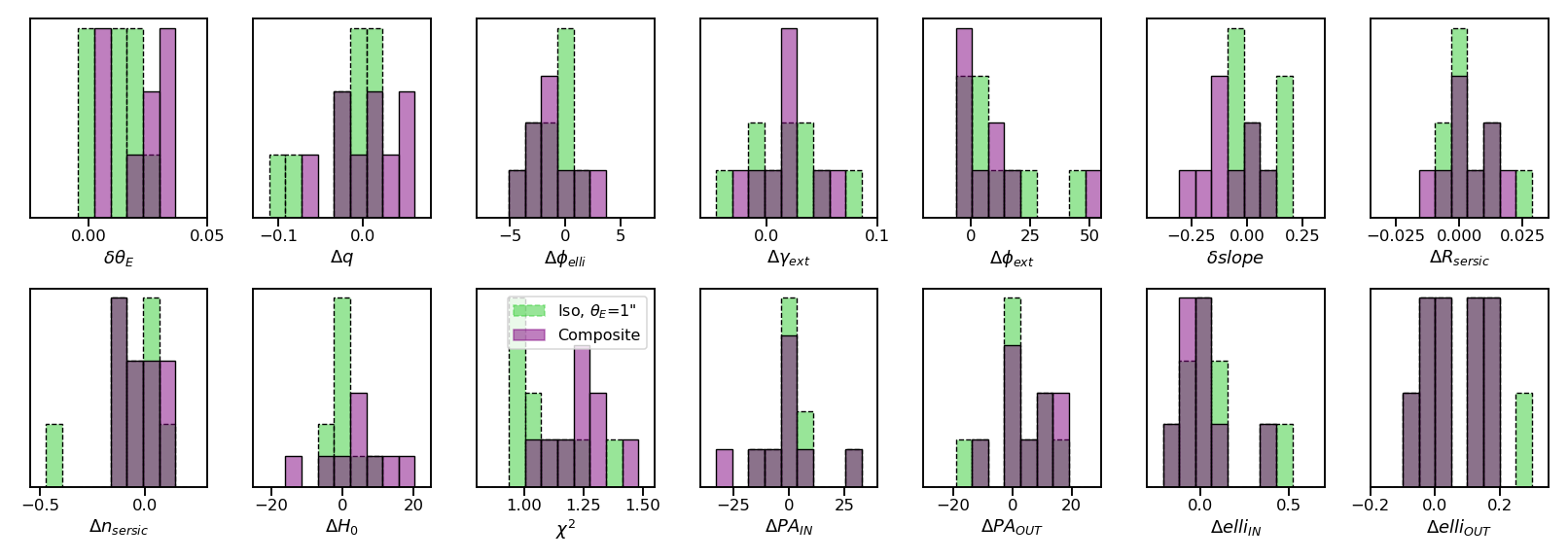}
    \caption{Comparison between fitted parameters results of mock images created with either isothermal (dashed green) or composite radial mass profiles (plain red). The mock lensing mass models display both ellipticity changes and twists based on the \textbf{hydro-simulated morphologies} using EAGLE galaxies from \cite{Mukherjee2018}. Only $\theta_E=1\arcsec$ lensing galaxies are displayed for the isothermal cases in order to match the Einstein radius of the composite mock images samples. Only fits with resulting $\chi^2<1.5$ are considered. Due to the $\chi^2$ cut, the displayed distributions of ellipticity gradient and twists are slightly different.}
    \label{fig_composite_eagle_results}
\end{figure*}

Fig. \ref{fig_composite_korm_results}-\ref{fig_composite_eagle_results} display the results when both twists and ellipticity changes are considered for the mocks based on either an SIE input mass profile and Einstein radius of 1$\arcsec$ (dashed green), or a composite input mass profile (plain red). Fig. \ref{fig_composite_korm_results} specifically focuses on results from mock lensing galaxies based the isophotes of the observed sample, and Fig. \ref{fig_composite_eagle_results} displays those based on the isodensity contours of the hydro-simulated sample (see Sect. \ref{sect_realistic_samples} for the sample details). Only fits with $\chi^2<1.5$ were considered. For the sample based on observed morphologies, all the fits have $\chi^2$ below the cutoff. This is not the case for the sample based on simulation: 9 (10) out of 12 mock images with a composite (isothermal) radial lens profile were fit to the required level. The galaxies leading to poor $\chi^2$ beyond the cutoff are not systematically the same for the two radial profiles (i.e., composite or isothermal). 

Compared to the models of the isothermal mock lensing systems, the parameters of the models resulting from composite mocks are more widely spread in terms of shear angle, slope, source size, and $H_0$ for azimuthal variations based on either an observational sample (Fig. \ref{fig_composite_korm_results}) or a hydro-simulated sample (Fig. \ref{fig_composite_eagle_results}). This extent in these characteristics is understood as the imprint of the variation in slope with angle for a given radius, as shown in Fig. \ref{fig_composite_azimuthal_slope}. Depending on the configuration and the ellipticity, the slope distributions even of mocks displaying neither twist nor ellipticity gradients are between twice and three times broader for the composite mass profile than for the isothermal profile. This scatter on the slope 
is additionally broadened by the ellipticity variations, hence the twice or three times larger scatter in $\Delta H_0$ for the composites mocks compared to the isothermal mocks. 

The interplay between the radial profile and the azimuthal variations does not add bias to $H_0$. Even if the MSD induces a shift in the central value of the $H_0$ distribution for the composite mocks -- the $H_0$ distribution for the composite mocks is centered on 60\,$\text{km}\,\text{s}^{-1}\,\text{Mpc}^{-1}$ instead of 70\,$\text{km}\,\text{s}^{-1}\,\text{Mpc}^{-1}$ (see \cite{TDCOSMOIV} for a discussion on how to mitigate this degeneracy)--, the $\Delta H_0$ distributions (see Table \ref{table_description_param} for definition) are centered on 0 for the composite mocks, similarly as for the isothermal mocks.

To summarize, the results of the modeling of composite radial mass profile are similar to those with an isothermal radial mass profile: the same correlations are observed, and no additional bias on $H_0$ is observed due to the interplay between radial and azimuthal variations. The only noticeable difference is the larger scatter of the distributions of the composite results. It is understood as the manifestation of an intrinsic property of the composite mass models independent of the azimuthal variations, however. The mock composite mass profile thus does not seem to interact more with the input azimuthal variations than a power-law mass profile does. 

\end{document}